\documentclass[screen]{acmart}

\AtBeginDocument{%
  }

\setcopyright{none}
\copyrightyear{2026}

\begin{document}

\title{Now You See Me: Designing  Responsible AI Dashboards for Early-Stage Health Innovation}

\author{Svitlana Surodina}
\email{svitlana.surodina@kcl.ac.uk}
\orcid{0000-0002-9444-0917}

\affiliation{%
  \institution{OPORA Health Technologies LTD}
  \city{London}
  \state{}
  \country{United Kingdom}
}

\author{Sinem G{\"o}r{\"u}c{\"u}  }
\author{Lili Golmohammadi}
\author{Emelia Delaney}
\author{Rita Borgo }

\affiliation{%
  \institution{King's College London}
  \city{London}
  \state{}
  \country{United Kingdom}
}

\renewcommand{\shortauthors}{Trovato et al.}
\begin{abstract}

Innovative HealthTech teams develop Artificial Intelligence (AI) systems in contexts where ethical expectations and organizational priorities must be balanced under severe resource constraints. While Responsible AI practices are expected to guide the design and evaluation of such systems, they frequently remain abstract or poorly aligned with the operational realities of early-stage innovation. At the ecosystem level, this misalignment disproportionately affects disadvantaged projects and founders thus limiting the diversity of problem-areas under consideration, solutions, stakeholder perspectives, and population datasets represented in AI-enabled healthcare systems.

Visualization provides a practical mechanism for supporting decision-making across the AI lifecycle. When developed via a rigorous and collaborative design process, structured on domain knowledge and designed around real-world constraints, visual interfaces can operate as effective sociotechnical governance artifacts enabling responsible decision-making. 

Grounded in innovation-oriented Human-Centered Computing methodologies, we synthesize insights from a series of design studies conducted via a longitudinal visualization research program, a case study centered on governance dashboard design in a translational setting, and a survey of a cohort of early-stage HealthTech startups. Based on these findings, we articulate design process implications for governance-oriented visualization systems: co-creation with stakeholders, alignment with organizational maturity and context, and support for heterogeneous roles and tasks among others. This work contributes actionable guidance for designing Responsible AI governance dashboards that support decision-making and accountability in early-stage health innovation, and suggests that ecosystem-level coordination can enable more scalable and diverse AI innovation in healthcare.

\end{abstract}


\maketitle

\section{Introduction}

Resource-limited AI innovation teams in HealthTech face a growing governance challenge. While regulatory and ethical expectations are high in the domain, organizational capacity to interpret, operationalize, evidence, and sustain compliance is limited.
Existing Responsible AI (RAI) tools are largely designed for mature organizations with established processes, leaving early-stage innovators without practical, context-aware support\cite{Sadek2024, Hollanek2025}. As a result, Responsible AI development becomes both burdensome and uncertain, slowing innovation in sensitive domains such as healthcare.
Responsible AI practices' translation from theory into an integral part of decision-making remains a major barrier: teams must interpret high-level ethical standards, diverse and changing regulations, generate evidence aligned with clinical practices, and manage multi-jurisdictional requirements, often without specialized governance expertise. Visual governance tools offer a promising way to mediate this complexity of regulatory and stakeholder requirements. In particular, Responsible AI governance dashboards simplify the contextual integration of insights into key decision points and daily practices, and offer more usability for non-technical, broader stakeholders. Rather than proposing a single “best” tool, the key need is a systematic, context-sensitive design approach for building governance dashboards that fit early-stage HealthTech workflows and can be adapted across organizations. Such a systematic approach enables ongoing innovation and advances both Visualization and RAI research in the rapidly evolving field while remaining grounded in real-world practice. 

AI-based innovation in healthcare is positioned to improve patient outcomes, reduce operational burden, and enhance productivity; yet real-world adoption depends on robust governance and repeatable practices that teams can operationalize \cite{England2024, DoH2025}. Innovation hubs, government funders, and translational programs increasingly act as intermediaries that support innovation, and there are expectations from such organizations to take a more proactive role in brokering informational, technological, infrastructural, and compliance resource access that would actively support AI projects \cite{PLN2022}. Recent research frames Responsible AI as an ecosystem-level challenge rather than a problem of individual models or organizations: an effective adaptation at national and international scale depends on shared assurance practices, intermediary institutions, and reusable governance infrastructure that can translate high-level principles into operational work under real constraints\cite{Alanoca2025}. Positioned within this agenda, our work treats governance visualization as shared infrastructure for capacity building and interpretive oversight. We adopt the following research questions: 1) How can governance dashboards be designed to make Responsible AI obligations operational for early-stage HealthTech teams, without requiring specialist compliance expertise? and 2) What mechanisms and constraints shape the process effectiveness? We use a two-stage visualization design framework and domain characterization principles, adapting them to Responsible AI dashboard design through the  findings of our research. Our contributions therefore are: 
\begin{enumerate}
    \item Empirical findings related to early-stage HealthTech AI governance needs and dashboard design-study approaches.
    \item A set of recommendations for designing Responsible AI governance dashboards that fit real-life innovation workflows.
    \item Pathways for operationalizing the design framework and resulting dashboard patterns as reusable, ecosystem-level infrastructure across innovation hubs and translational programs.
\end{enumerate}

We conclude that governance dashboard design is most effective when it balances resources with obligations, is rooted in innovation lifecycle realities, staging obligations relative to technical maturity,  supporting interpretation across heterogeneous jurisdictions, enabling co-production among stakeholders using multi-stage, collaborative design study frameworks, and generating reusable governance artifacts that integrate into stakeholder incentives and workflows.

The insights will be useful to designers and HCI researchers building Responsible AI tools, applied RAI governance researchers, and innovation hubs, accelerators, and translational program and policy leads supporting early-stage health AI innovation.

\section{Related Work}

\subsection{Responsible AI Governance in Healthcare}

Responsible AI literature increasingly frames governance in healthcare as a sociotechnical phenomenon, a clinical and institutional challenge rather than a purely technical problem. AI systems in health settings operate within dense and complex environments shaped by clinical risk, professional responsibility, regulatory oversight, and ethical obligation. Failures of governance frequently stem from 'wicked problems' such as misabstraction where design practices struggle to incorporate organizational contexts that structure clinical decision-making \cite{deTroya2025,Sabuncuoglu2025}. In this view, governance cannot be reduced to translating abstract principles into technical controls; practices co-evolve through situated, non-linear processes embedded in real-world workflows \cite{Ruster2025,Sadek2024}. Empirical studies of healthcare industry and research further show that access to regulatory expertise and resources fundamentally shape what forms of Responsible AI governance are feasible in practice \cite{Scheuerman2024,De2025}. Assurance mechanisms aimed at operationalization of governance include, \textit{inter alia}, internal and external auditing frameworks that position accountability as a longitudinal property that depends on traceable evidence spanning data curation, model development, deployment, and post-deployment use \cite{Raji2020,Panigutti2025}, documentation practices, including dataset and process-level artifacts \cite{Pushkarna2022,Lucchesi2022,Mueller2020}. More recent approaches extend these mechanisms through argument-based assurance, emphasizing justified evidence collection and dynamic monitoring rather than static compliance artifacts \cite{Sabuncuoglu2025}. However at present, systematic reviews of algorithm auditing show that existing practices remain geographically, linguistically, and demographically skewed, raising concerns about which clinical contexts, populations, and harms are rendered visible through prevailing assurance regimes \cite{Urman2025}.

These challenges are modified and intensified by the transition from voluntary guidance to binding regulation in healthcare AI. Comparative analyses of global AI regulation demonstrate substantial divergence in scope and enforcement, increasing interpretive burden for healthcare innovators operating across jurisdictions \cite{Alanoca2025,EUAIAct}. Within the European context, the EU AI Act is noted for robustness, cybersecurity, and risk management as governance priorities that intersect with fairness and transparency concerns \cite{nolte2025,Steimers2022,Baquero2020}. At the same time, adversarial analyses of regulatory avoidance show that risk-based AI regulation incentivise legally compliant but strategically unsustainable avoidance strategies, and calls for more supportive interpretive guidance and tools for treating compliance as an upstream design activity \cite{Yew2025}. AI governance in healthcare must account for broader socio-economic, environmental, and structural dimensions of harm \cite{Roy2025,Kinney2025}. Intersectional and ecofeminist critiques are fully applicable to this domain: tools and policies risk reproducing existing power relations if they abstract away from social context, sustainability, and lived experience \cite{Vethman2025,Valdivia2025}. 

Human-centered perspectives emphasize stakeholder involvement across the AI lifecycle to address bias and ensure equitable outcomes \cite{Chen2023}. For instance, critical evaluations of LLMs in mental health contexts demonstrate that explainability and safety mechanisms remain insufficient for high-stakes substitution of human expertise \cite{Moore2025}. 

A recurring limitation of the state of art research is the assumption of organizational capacity. Policy guidance and industry frameworks often frame Responsible AI as a set of best practices without addressing how small or early-stage healthcare teams can sustain documentation and monitoring over time \cite{AWSAI,Accenture,McKinsey,LovelaceInstitute}. Healthcare-focused reviews consistently identify gaps between ethical guidance and operational reality, particularly in early-stage or resource-constrained settings where clinical, regulatory, operational, and commercial pressures intersect \cite{Morley2020,Siala2022,Gerke2020,Rigby2019}. Further, analyses of AI supply chains and public procurement in health reveal that accountability is distributed across stakeholders across ecosystems, challenging single-project-centric models of governance \cite{Hopkins2025,Johnson2025}.

A gap remains to be addressed in the systematic approach to design of practical, human-centered tools that can put Responsible AI in practice within the constrained workflows of AI innovation.

\subsection{Visualization and Dashboards}

Visualization design provides a methodological pathway for operationalizing Responsible AI governance by translating sociotechnical contexts and demands into interpretable, situated representations aligned with the lived realities of early-stage healthcare teams. Following established visualization terminology, we treat a dashboard as a visual output that consolidates the most important information needed to achieve one or more objectives. In this paper, a governance dashboard is such a  visual artifact supporting oversight and decision-making across heterogeneous users. It can serve as an instrument for engaging wider non-technical stakeholders by translating model-, data-, and compliance-related information into role-relevant views, enabling busy or resource-limited participants (e.g., clinicians, oversight board members, even patients) to stay in the loop without needing direct access to operations. 

Domain literature positions visual analytics as a scaffold for integrating AI into clinical decision-making \cite{Prince2025, Greenes2018}. Human-Computer Interaction studies of decision-making demonstrate that visual interfaces improve decision quality and speed \cite{Oral2023,Zhuang2022,Neri2025}; dashboards have a growing role in healthcare policy and crisis response and provide design space for balancing clarity, responsiveness, and contextual information \cite{Schulze2023}; deployed systems demonstrate that dashboards become embedded in the safety-critical workflows and inherit governance responsibilities related to monitoring and documentation \cite{Wang2023}. At the same time, implementation studies consistently show that adoption is constrained by organizational readiness rather than technical performance alone \cite{CTS2022,PLN2022}. Visually sophisticated systems can fail when they rely on expertise or maintenance labor that organizations cannot sustain \cite{Alpherts2024}. Inclusive visualization design is critical: differences in age, skill, and ability, when neglected, can turn visualization from a decision support tool into a barrier that excludes stakeholders from governance and accountability work \cite{Kim2021, Lundgard2019}.

A body of research delivers on diverse visualization approaches that explicitly support governance-relevant practices. Visualizations of data preprocessing decisions make upstream choices inspectable and accountable \cite{Lucchesi2022}. Dashboards for equitable work allocation demonstrate how interfaces can surface responsibility and negotiation rather than obscuring them \cite{Liang2025}. Antweiler et al. \cite{Antweiler2022} demonstrate how visualization can support expert inspection and accountability in healthcare prior to AI deployment. Standards for ethical considerations in emulated empathy further guide interface design choices to carry normative weight in sensitive domains \cite{IEEE-standard2024}. Field-level syntheses call for greater attention to provenance, uncertainty, and evaluation rigor in real-world visual analytics applications \cite{Wu2023}. Based on the variability and evolving nature of the field, it emerges that governance-oriented dashboards cannot be designed as generic artifacts, but require systematic and efficient in-situ design approaches that account for evolving governance needs and tasks, motivating the design study methodologies described next.

\subsection{Design Study Methodologies and Domain Characterization for Responsible AI Dashboards}

Visualization design study methodologies provide a well-established foundation for developing systems in complex, real-world settings. Seminal frameworks emphasize deep domain immersion, iterative co-design with stakeholders, and reflective abstraction as core methodological precepts \cite{SMM2012,MUN2009,munzner2014visualization}. These approaches position visualization systems as interventions that evolve through sustained engagement with domain practices, constraints, user contexts, and institutional realities. In the safety-critical and highly regulated domain that is healthcare, design studies further highlight the need to account for stakeholder networks with asymmetry of authority, expertise, and burden of long-term impact \cite{Hall2020,Marai2018}.

Recent methodological work has increasingly problematized the sustainability of design studies in governance-heavy contexts:  information needs and other factors change over time and across roles, limiting the effectiveness of one-off design and evaluation cycles \cite{Turri2024,Ye2020}. Meta-analyses of visualization applications call for reusable abstractions and shared infrastructure to support longitudinal impact beyond individual case studies \cite{Wu2023,Xing2024}. Within this literature, domain characterization emerges as a critical methodological asset. Rather than treating domain understanding as a project-specific precursor, domain characterization stabilizes knowledge about tasks, stakeholders, organizational structures, and constraints that recur across projects, enabling reuse and comparison \cite{Marai2018,Cibulski2022}. This line of work has been extended through industry-level domain characterization frameworks developed for AI-enabled clinical decision support systems \cite{Surodina2024,Surodina2025}. Design studies in AI-enabled clinical contexts demonstrate that ethical and regulatory considerations cannot be retrofitted at later stages but must be embedded upstream through domain characterization and iterative stakeholder engagement \cite{MMA2014,Preim2020,Preim2018}. 

For Responsible AI dashboard design, these methodological advances are particularly relevant and motivate design study approaches that couple domain characterization with dashboard design for Responsible AI, treating governance dashboards as interfaces built on structured domain knowledge rather than standalone visual artifacts. Grounding design in reusable domain characterizations enables visualization research to better support early-stage HealthTech innovation, where teams must navigate complex governance demands without the organizational capacity assumed by much of the existing literature.

In summary, across Responsible AI research, there is distinct lack of work on empirically rooted governance-enabling visualization design approaches, while visualization work in healthcare rarely focuses on the problematics of RAI for early-stage innovation as first-order design concern. 

\section{The Framework and the Longitudinal Program}
\label{Framework}

\begin{figure}[h]
  \centering
  \includegraphics[width=\linewidth]{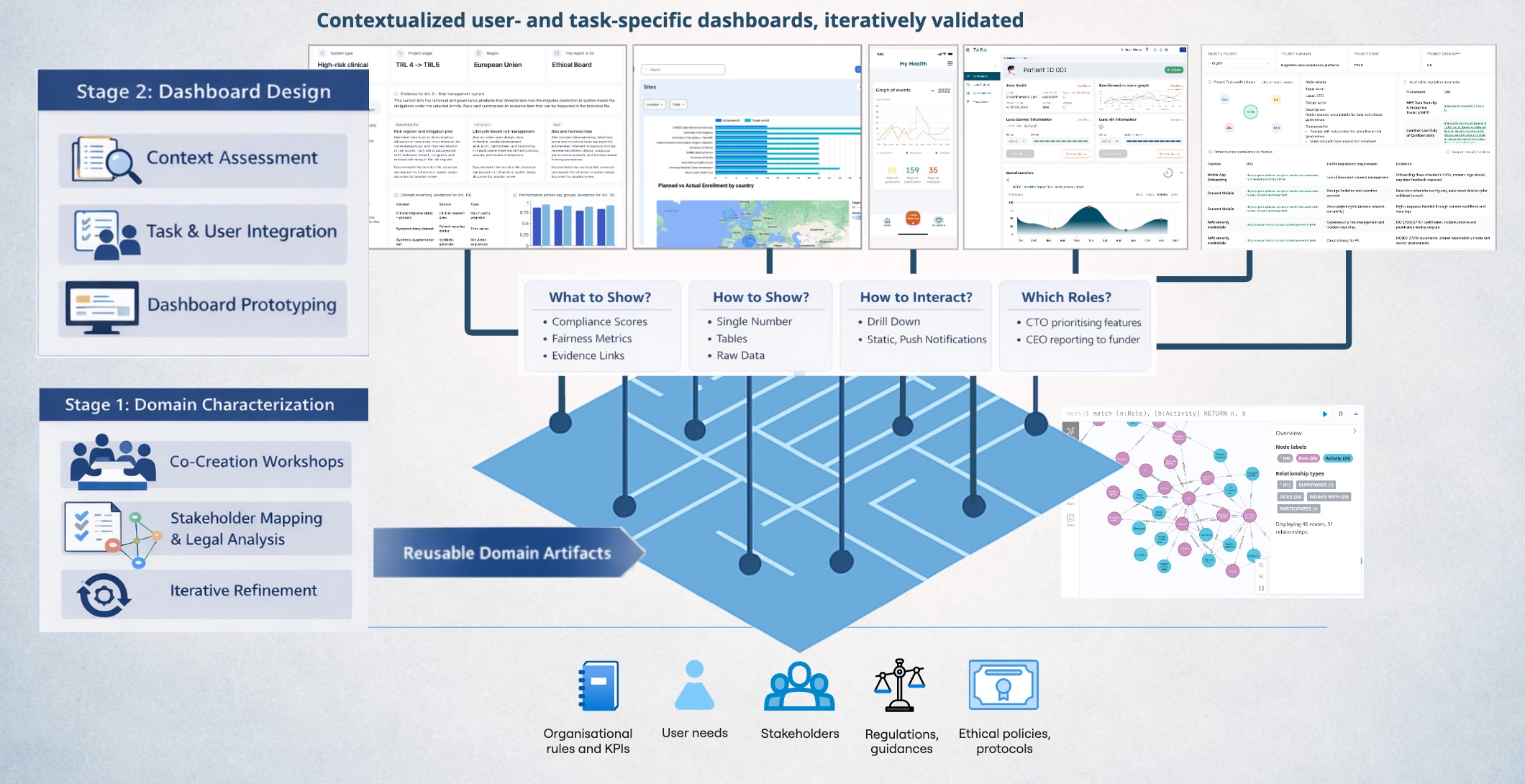}

  \caption{ Overview of the two-stage visualization design-study approach: co-created industry characterization (Stage 1) produces reusable domain artifacts that surface stage-specific obligations, evidence dependencies, stakeholder responsibilities etc., which then define and constrain dashboard design (what is shown, how it is encoded, and how roles interact) (Stage 2) and support context- and task-specific views for users-decision-makers. }
  \Description{}
    \label{ds}
\end{figure}

This paper draws on a multi-year longitudinal research program, within which the present study reports a focused analysis. This research program is theoretically grounded in a two-stage visualization design study methodology \cite{Surodina2025} and participatory practices \cite{Janicke2020,Isenberg2011,Kerzner2020,Akbaba2023,Hall2020}. The two-stage design framework distinguishes between reusable, industry-level domain understanding and project-specific design. In the first stage, an industry characterization is constructed to externalize stable preconditions that shape work across organizations, including regulatory obligations, ethical constraints, stakeholder roles, organizational capacities, data governance requirements. These elements are synthesized from literature, expert engagement, and documentary analysis via an industry-level data collection program centered on purposive sampling across heterogeneous healthcare-AI stakeholder roles (technical teams, clinicians, data stewards, IT managers, administrators, regulatory experts, ethicists); data were gathered through semi-structured interviews and a focus group using case-based probes, with sessions recorded (with consent) and transcribed for analysis with Otter.ai \cite{Otterai2025}. It triangulated interview findings with document analysis of institutional policies, national healthcare IT strategies, data governance frameworks, and regulatory guidance (e.g., GDPR, EU AI Act, HIPAA, FDA, MHRA), then applied thematic analysis by coding transcripts and documents into predefined domain-parameter categories, updating the parameter table when new themes emerged. The outputs served as inputs to an industry characterization knowledge-graph tool built in Neo4j (\cite{Neo4j2025}), supporting visual representation and analytical capabilities. The second stage used this characterization for project-level visualization design, enabling individual design studies to focus on situated tasks and workflows without repeatedly rediscovering structural constraints. The two-stage methodology is illustrated by Figure \ref{ds}. The outputs were validated through member-checking with a subset of experts, and treated as a living artifact intended for reuse and revision as regulatory and technical conditions shift. The framework and the toolbox had been used to implement real-world projects.  

The research is embedded at the intersection of academia and industry; ongoing collaboration with HealthTech AI innovation projects via a partner digital health technology innovation consultancy provides access to decision-making processes and constraints in situ. This approach has both advantages and disadvantages: while it strengthens systematic applicability and enables iterative, practice-grounded refinement, it can also bias problem framing toward partners’ priorities and introduces selection bias. This was addressed by validating findings that were consistent across projects and triangulated across several evidence streams. The design-study material was collected from HealthTech organizations spanning Technology Readiness Levels (TRL) \cite{EuropeanCommission2017} 1-7 and multiple European jurisdictions. Data include onboarding and follow-up interviews, co-creation workshops, task tracking information in Jira \cite{Jira2025}, technology team meeting logs, and documentary artifacts (including reporting templates and submitted reports, governance library), and a Knowledge Graph (KG)-structured database of reusable project assets. 

\begin{figure}[h]
  \centering
  \includegraphics[width=1\textwidth]{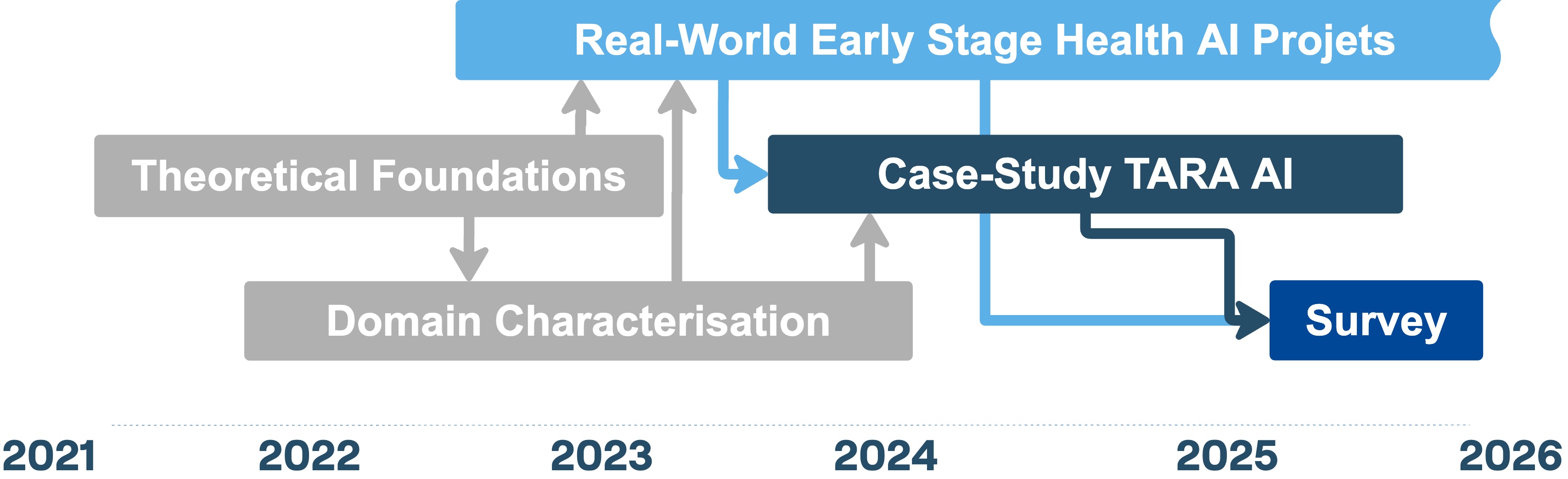}

  \caption{ Project timeliness and scope: gray denotes past activities, while blue shades cover the base study learnings reported here pertinent to the design of RAI dashboards. First, learnings are synthesized from real-world projects and informed the Case-Study, and both subsequently led to the formulation of Survey questions.}
  \Description{}
    \label{time}
\end{figure}

The overall program is a longitudinal research effort initiated in 2021 and continuing through 2025 and beyond. Within it, during 2023-2025, we collected data on responsible AI requirements and explored the role of governance dashboards in AI innovation
(Figure \ref{time}). 

\section{Methods}

The empirical basis of the present study comprises three components.

\begin{figure}[t]
  \centering
  \begin{minipage}{0.49\textwidth}
    \centering
    \includegraphics[width=\linewidth]{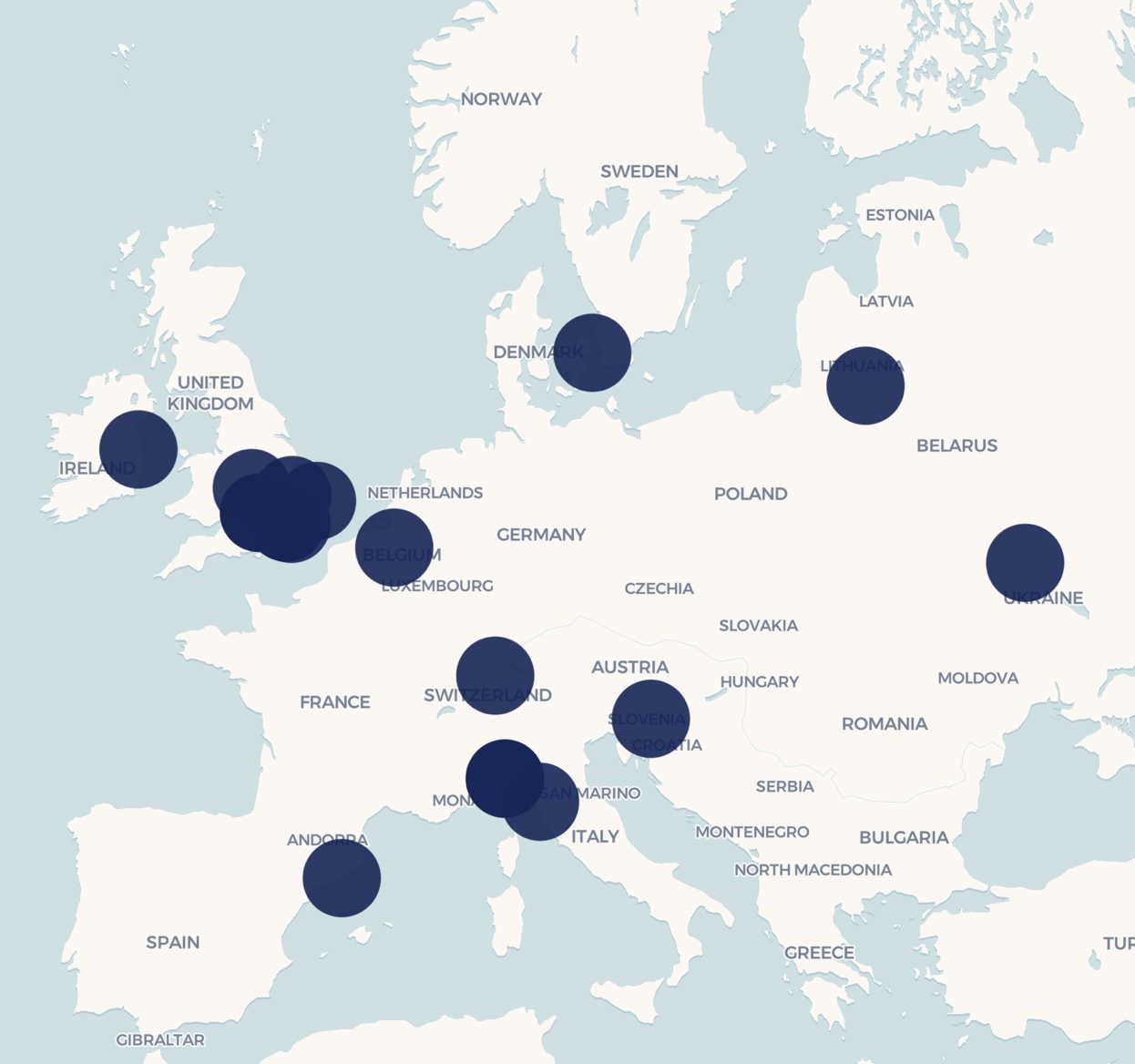}
    \caption{
Projects span European jurisdictions with different AI regulatory regimes (incl. the UK, EU, EEA, Ukraine), with 43 \% of teams women-led, and several solutions focused on or using datasets from underserved populations, such as workers in care homes.}
    \label{map}
  \end{minipage}\hfill
  \begin{minipage}{0.49\textwidth}
    \centering
    \includegraphics[width=\linewidth]{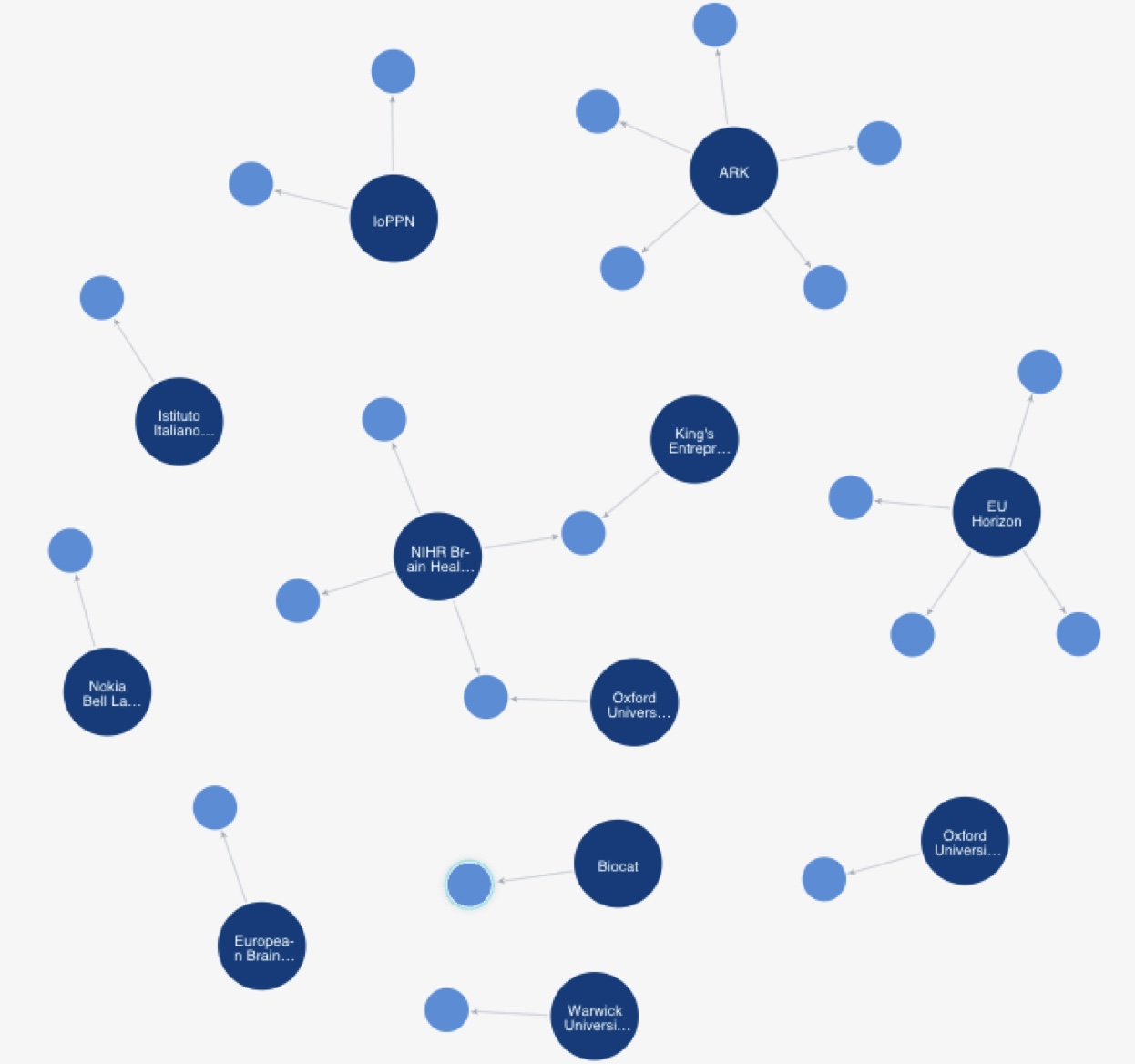}
    \caption{The 21 participating organizations were each connected to one or more innovation clusters through funders, spin-outs, or accelerator programs; these ecosystem-level linkages were not planned a priori and emerged through analysis at a later stage.}
    \label{org}
  \end{minipage}
\end{figure}

\textbf{Real-World Projects: Teams Lived Experiences}. Building on the program described in Section \ref{Framework}, we synthesize insights from longitudinal design studies with 21 HealthTech projects (Figures~\ref{map} and~\ref{org}) at different stages of implementation: TRL 1-6 and varied correspondingly in organizational structure and maturity. The participating organizations developed or deployed AI-based solutions ranging from systems where AI constituted a core component (e.g., machine learning or retrieval-augmented generation) to AI-enabled tools for clinical and consumer decision support. 
Data were collected through embedded collaboration between 2021-2025, with Responsible AI governance as an explicit analytic focus during 2023-2025.

 We aggregated qualitative and operational traces from (a) 84 recorded call transcripts (including onboarding project team interviews, project coordination, planning, compliance discussions, and product meetings), (b) contemporaneous project notes and decision logs, and (c) task backlogs and issue trackers. To reduce gaps in the historical record and clarify governance-related decision points, we additionally conducted open-ended interviews with project managers (focused on constraints shaping feasibility, tensions). Where available, we also collected artifacts used for external accountability (e.g., funder application forms, reporting templates, ethics submissions, and certification planning documents) to triangulate how governance expectations corresponded to internal work. We coded materials for triggers of Responsible AI activity (e.g., funding, ethics review, certification, cross-border expansion); recurring conflicts and uncertainty points; capacity constraints and workarounds; and moments where visualization or structured representations were introduced, developed, tested. This analysis produced a cross-project narrative of recurring design requirements and processes that informed subsequent dashboard instantiation in the case study and the hub-cohort survey.

\textbf{Case-Study: Governance Dashboard for Migraine Decision Support.} Second, we conducted a demonstrative case-study of an EU-funded AI-enabled clinical decision support system for migraine prediction and management in resource-constrained settings to illustrate how the two-stage framework and associated design process principles and learnings from the previous step are operationalized in practice. The project operated within the European regulatory context, using the EU AI Act as the primary reference for Responsible AI obligations. Key stakeholder groups included clinicians, a formally present but in practice under-deployed ethics oversight board, a technical delivery team, EU auditors, and a project lead responsible for coordination and reporting. Accordingly, regulatory compliance, evidence generation, and clinically grounded explainability aligned with ethics requirements were central design constraints across the project lifecycle. The AI implementation began in 2023 and progressed through initial modeling, ethics approval, EU validation planning, real-world data collection, evaluation, and deployment with patients; ethics board approval was obtained for data collection and system deployment.

\textbf{Survey: Governance in Early-Stage HealthTech}. Third, to map the pathways for scalability, probe the limits of organization-centric governance and assess appetite for shared infrastructure, we ran a survey with 16 early-stage HealthTech startups (TRL 3-6) drawn from one accelerator cohort. The cohort sampling is intentionally homogeneous in sector and geography to support comparison under similar constraints. The survey combines closed-response items (readiness, openness to shared tools) and open-ended prompts (needs, concerns, affect). All responses were collected on a single day, in a separate moderated session per startup: the CEO or a co-founder of each project completed the questionnaire live while a researcher probed for clarification and concrete examples, producing both the recorded survey responses and contemporaneous notes used in analysis. The completed paper forms and notes were transcribed next day into a digital system for aggregation and analysis.

\section{Results}
\subsection{Real-World Projects: Teams Lived Experiences}

During the onboarding interviews, none of the projects explicitly articulated Responsible AI factors as design requirements beyond regulatory compliance or certification needs. When discussed, terms such as “Responsible AI” and “Ethical AI” were used interchangeably and rarely connected to concrete organizational priorities or workflows. At the same time, regulatory compliance, grant requirements referencing Responsible AI, ethics board approvals, and certification pathways consistently ranked among the most important concerns for founders and project leads. Prioritization across projects was strongly shaped by immediate goals related to product and technology validation, fundraising, and clinical evidence generation, and often treated as sequential, not parallel processes. As one CEO commented: \textit{“There’s the big regulatory journey that we have to do, and we’re not able to do that unless we’ve got money to do it. We first have to do an implementation study… publish the clinical effectiveness data… then the compliance journey.”}

Product development activities and milestones were tightly coupled to funding rounds, with funders often shaping how compliance efforts were defined and pursued. Several teams reported openness to switching countries or regulatory pathways depending on which fundraising efforts come to fruition.

Team size ranged from single founders or principal investigators to groups of 10-15. Some organizations had formally designated clinical safety or data security responsibility, but such roles were mostly symbolic and did not translate into sustained governance practices. None of the projects had explicit Responsible AI roles or ownership defined. In later-stage organizations, engagement with third-party compliance providers became more common, but this was typically reactive rather than integrated into core design processes.

Across 18 of the 21 organizations, regulatory and ethical obligations were described as central to funding, certification, and clinical validation decisions, yet teams reported limited understanding of which obligations applied at a given stage, what constituted sufficient evidence, or how governance activities aligned with product and organizational milestones. By contrast, technical development tasks were perceived as challenging but legible and actionable. Compliance was repeatedly characterized as a “big” and “scary” uncertainty and risk, in stark contrast to teams’ confidence in their capabilities to lead technical delivery. 

These gaps manifested concretely in design and development trade-offs. When prioritizing features in development backlogs, governance-related concerns such as documentation or evidence tracking were consistently deprioritized in favor of faster product testing. Technical teams often treated compliance as a separate feature or downstream activity, rather than as an integrated aspect of system design and development. Introducing visualization designs that explicitly linked regulatory requirements to system features and surfaced them at key workflow points, such as design reviews and development prioritization meetings, helped reconnect governance priorities with the AI development lifecycle  (Figure~\ref{ds}). Such shifts made previously implicit Responsible AI considerations visible within everyday decision-making processes. When surfaced, the regulatory considerations became discussed more often in cross-team meetings, however quickly deteriorated in utility and impact unless closely linked to the operational priorities and stakeholders' tasks.

\subsection{Case-Study: Governance Dashboard for Migraine Decision Support (TARA AI)}

\begin{figure}[h]
  \centering
  \includegraphics[width=\linewidth]{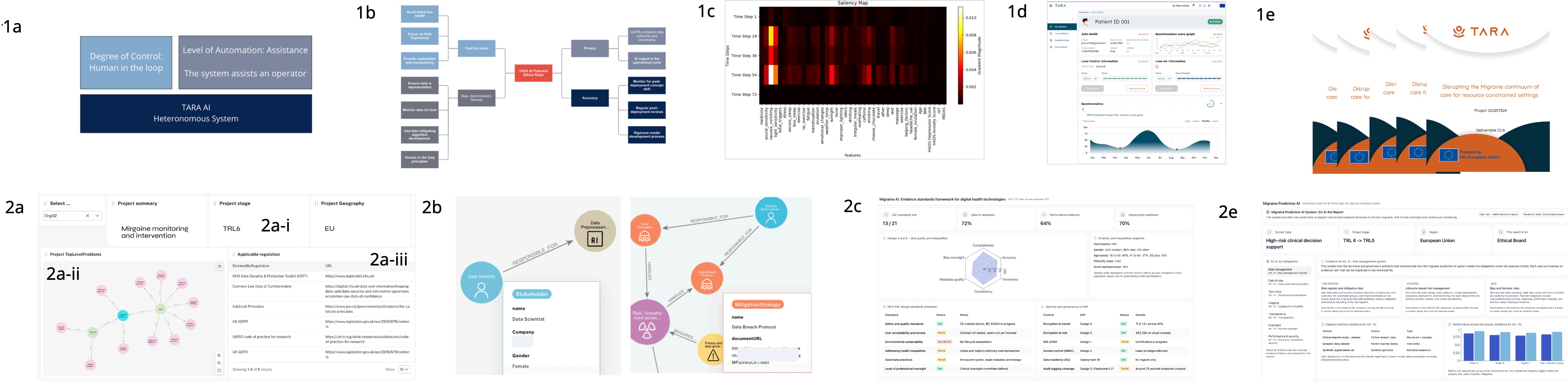}
  \caption{Case study overview: operationalizing Responsible AI through the governance dashboard design. Top row, before (1a–1e): the migraine decision-support team relied on fragmented, manual governance work. System characterization and risk framing were handled through ad hoc documentation and informal interpretation (1a–1b), while technical model-facing artifacts (for example, model saliency maps) were difficult for non-technical or time-constrained stakeholders to interpret and use in decision-making (1c). Reporting was assembled post hoc across multiple tools and formats  (1d), producing slow, reviewer-heavy evidence packages, cumbersome reporting (1e) and limited internal alignment on what counted as “sufficient” Responsible AI evidence at a given stage.
Bottom row (2a–2e, “after”): using the two-stage Sustainable Design Study framework and industry precondition mapping, the team rapidly configures governance views by TRL and target jurisdiction (2a), grounded in explicit stakeholder roles, obligations, and evidence dependencies represented as structured domain artifacts (2a-ii–2b). This enabled project-level governance dashboards that connect regulatory obligations to concrete evidence links and responsibility allocation for different users (2c and 2e) and ongoing external reporting enabling faster feedback.}
  \Description{}
  \label{raicase}
\end{figure}

At the outset, the project faced substantial uncertainty regarding applicable regulatory
requirements and evidence expectations, Figure \ref{raicase}. Existing regulatory documents and supporting guidance from the EU was experienced as
fragmented and checklist-oriented, with problematic mapping to the project’s technical and clinical context. The team lacked dedicated compliance personnel or budget, requiring
governance tasks to be distributed across existing roles. While a range of technical and clinical artifacts had already been produced, these assets were not organized or linked
in ways that supported regulatory reasoning or external review. Despite this uncertainty,
governance work needed to begin early to support prospective design decisions, clinical validation planning, and preparation for CE-Mark certification.

Using the existing industry characterization and knowledge-graph-based governance model as
infrastructure, a governance dashboard was designed through iterative configuration and refinement as a visual layer translating
regulatory texts into interpretable views tailored to different stakeholders. Governance
activities were explicitly linked to innovation milestones (clinical study
design, grant reporting, and regulatory planning). Requirements and other parameters predefined in the characterization graph were refined through interviews with
project stakeholders and mapped to evidence requirements already encoded in the graphs. This enabled existing technical artifacts to be linked as evidence without
restructuring development workflows, supporting rapid configuration under time constraints.

Multiple dashboard iterations were developed to explore alternative representations of
obligations and dependencies. Clinician-facing views were designed to integrate with patient assessment
workflows, prioritizing data quality, provenance, and explainability of migraine predictions, such as associations with sleep patterns. Ethics and policy stakeholders used
the dashboard to examine regulatory coverage, however score-based representations were not evaluated highly.

Interpretive analysis of the embedded design and implementation is based on the insight gathered in stakeholder interviews, informal discussions, and observation of use during ongoing project work. 

While the case illustrates how the design-study process can be enacted within a single project, it also surfaced limits to organization-centric
approaches. Several governance activities remained structurally difficult to sustain at
the project level, motivating questions of reuse, scalability, and shared infrastructure.
These observations informed the design of a survey of innovation
teams within a single innovation hub cohort enabling examination of governance capacity
and openness to shared, ecosystem-level support under comparable conditions.

\subsection{Survey: Governance in Early-Stage HealthTech}

\begin{figure}
    \centering
\includegraphics[width=\linewidth]{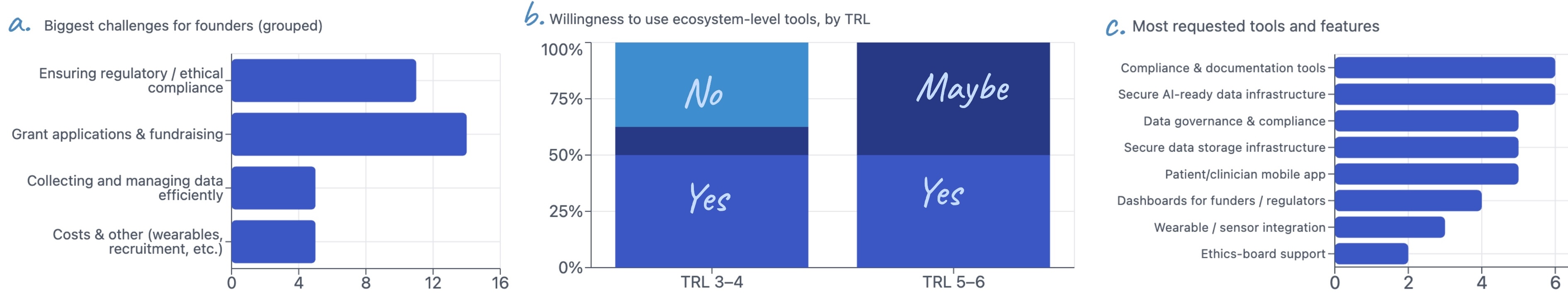}
    \caption{Survey results relevant to challenges and tooling: a) Compliance and fundraising dominate as the biggest challenges; b) Willingness to use ecosystem-level tools, by TRL. Hard “No” responses are concentrated in TRL 3–4. By TRL 5–6, teams split between Yes and Maybe; rejection disappears; c) Most requested tools and features in shared infrastructure: Founders prioritize compliance, secure AI-ready storage, with dashboards close behind.}
    \label{survey-three}

\end{figure}

Our thematic analysis produced several consistent quantitative and qualitative patterns:

\begin{itemize}
    \item Regulatory and ethical compliance was cited as the dominant challenge, exceeding
    concerns related to technical development (Figure \ref{survey-three}-a).
        \item Willingness to engage with governance tooling increased substantially from  TRL 3-4 to TRL 5-6, indicating a maturity threshold rather than principled
    resistance (Figure \ref{survey-three}-b).
    \item Strong demand was reported for compliance documentation, AI-ready
    data infrastructure, and governance support tools (Figure \ref{survey-three}-c).
    \item 81\% of startups reported openness to using shared infrastructure for
    data and governance.
    \item 12 out of 16 expressed willingness to undergo an AI readiness or governance
    assessment.
\end{itemize}

The patterns point to the fact that early reluctance toward Responsible AI tooling reflects capacity and timing constraints rather than rejection of governance goals.

Open-ended responses and follow-up comments revealed four recurring themes:
\begin{enumerate}
    \item Timing sensitivity: early-stage teams dismissed 'premature' governance work.
    \item Confidence gaps: founders reported uncertainty in interpreting
    regulatory and ethical requirements.
    \item Missing internal tools: governance tasks were handled manually or
    informally.
    \item Need for interpretability: teams sought governance support that
    explained obligations rather than presenting static checklists.
\end{enumerate}

Across responses, the emotional climate (''\textit{In your current stage of the startup journey, how would you describe your emotions?}'') can consistently be described as
\textit{''excited but stressed''}: the free-text field distinctly featured two sets of keywords: ''excited, hopeful, 
motivated'' and ''overwhelmed, uncertain,
tired, stressed'', very often at the same time.


\section{Discussion}

Our results address two questions: how the design process of governance dashboards can make Responsible AI obligations operational for early-stage HealthTech teams, and what mechanisms and constraints shape whether that effort works in practice. Across three complementary evidence streams (longitudinal design studies with 21 projects, the TARA-AI case, and a survey of 16 startups), a consistent pattern emerged: governance work was decision-critical yet difficult to interpret and implement within resource-limited workflows.  Teams treated regulatory and ethical obligations as decision-critical for funding, certification, and clinical validation, yet could not link which obligations applied at a given stage, what evidence would be sufficient, or how governance work mapped onto product milestones. In contrast, technical work was consistently described as difficult but tractable. This asymmetry suggests that the design objective is less about increasing motivation for RAI and more about converting ambiguity into staged, role-relevant work that fits existing work patterns.

The structured design approaches to dashboards make RAI compliance more operational in practice. First, staging obligations relative to maturity reduced overload and made governance feel actionable. In both the longitudinal design-studies and the survey, reluctance to engage with governance tooling concentrated the earlier stages of the projects and largely disappeared by TRL 5-6, consistent with a capacity threshold. Second, linking obligations to evidence dependencies supported reuse of material that teams already generated for technical delivery. In the TARA AI case, mapping EU AI Act oriented requirements to existing technical and clinical artifacts allowed evidence to be assembled and developed without creating a parallel compliance workflow. Third, accountability became more legible when views were role-specific and anchored to concrete decision points, such as clinical study design, grant reporting, and regulatory planning, rather than presented as generalized checklists or generic scores. This aligns with the case observation that score-based representations were not valued, while staged, interpretable readiness framing was. Finally, a structured, two-stage design approach allowed for more efficiency. Dashboards are effective for early-stage teams when they function as translation layers: from regulatory texts to stage-appropriate tasks, from tasks to evidence links, and from evidence links to role-specific responsibility allocation, all expressed in formats that can be used in everyday prioritization and external reporting.

\textbf{Mechanisms and constraints shaping effectiveness}. The results suggest four constraints that repeatedly limited effectiveness, even when teams recognized governance as important. (1) \textit{Capacity and timing}: governance work competed with fundraising and validation milestones, and was often deferred until forced by an external trigger. (2) \textit{Interpretive uncertainty}: guidance was experienced as fragmented and difficult to map onto a specific clinical and technical context, producing a persistent fear of doing the wrong work at the wrong time. (3) \textit{Workflow separation}: teams tended to conceptualize compliance as downstream or as a separate feature, which produced systematic deprioritization in backlogs. (4) \textit{Distributed responsibility without ownership}: dedicated governance roles were rare; even when nominal responsibility existed, it did not translate into sustained practice.

Governance dashboards function primarily as interpretive infrastructure that restructures decision-making rather than automating compliance. In the longitudinal projects, making requirements visible at key workflow points, such as design reviews and prioritization meetings, helped reconnect governance concerns to the development cycle. In the TARA AI case, the two-stage approach reduced set-up cost by starting from reusable domain artifacts and then refining them through stakeholder interviews, which supported rapid configuration under time constraints. The survey reinforces that these mechanisms match expressed needs: teams most frequently requested compliance documentation support, AI-ready data infrastructure, and governance tools, and a large majority reported openness to shared infrastructure and to undergoing an AI readiness or governance assessment.

We formulate the key design process implications below and give a list of recommendations in Table  \ref{table}:

\textbf{Design for staging and reuse.} Dashboards should surface what to do next at a given stage and how to reuse existing artifacts as evidence, because early-stage teams systematically deprioritize governance work that requires creating new documentation streams.

\begin{itemize}
    \item Design for interpretability over scores. The case indicates that score-based, overgeneralized representations. Interfaces should foreground traceable links between obligations, rationale, and evidence rather than collapsing interpretation into a single number.
    \item Design around triggers and decision points. In practice, governance work activates around funding, ethics review, certification, and cross-border expansion. Dashboards are more likely to be used when they are explicitly integrated into these decision moments and their surrounding artifacts.
    \item Design for role-specific accountability. Projects distributed governance tasks across existing roles. Dashboards should therefore clarify who needs to act, what evidence each role controls, and how responsibilities connect across stakeholders.
\end{itemize}

\subsection{From project-level dashboards to hub-level governance infrastructure}
The longitudinal material and the case both show limits to organization-centric governance: even with tailored views, keeping pace with evolving requirements and maintaining audit-ready evidence packages strained small teams. The survey strengthens the inference that shared infrastructure is a plausible complement, not because teams want to offload responsibility, but because they lack the stable capacity assumed by many Responsible AI frameworks.

\begin{table*}[h]
\centering
\begin{tabular}{p{3.2cm} p{8.5cm} p{5.5cm}}
\toprule
\textbf{Category} 
& \textbf{Recommendation} 
& \textbf{Technical Notes / Tools / Artifacts} \\
\midrule

\textbf{Domain Characterization} 
& Use structured mapping of stakeholders and regulatory obligations to ensure visualizations are designed according to and reflect real organizational dependencies. 

& Industry Characterization KG. \\

\textbf{Regulatory Encoding} 
& Encoding of regulations as (machine-actionable) rules to simplify action must be linked with a project stage, stakeholder capacity, traceability, geographical and other applicability.
Stakeholders need correct, interpretable guidance without legal expertise, and the ability to understand why a requirement applies. 
& Knowledge-graph rules; regulatory nodes and edges; rule-based inference; mapping tables. \\

\textbf{Workflow Alignment} 
& Stage requirements by TRL and organizational maturity to avoid overload and increase adoption. 
Stakeholders require dashboards that match their daily work, show only relevant obligations, and support realistic next steps. 
& TRL-based dashboard templates; configurable components. \\

\textbf{Interpretability \& Uncertainty} 
& Surface uncertainty and provide feedback loop to avoid false certainty. 
Stakeholders need transparency about ambiguous tasks and the flexibility to express disagreement. 
& Evidence completeness indicators. \\

\textbf{Visual Encoding Strategy} 
& Provide concise, role-specific views with actionable next steps. 

& Modular dashboard panels; compact progress views; next-steps. \\

\textbf{Evidence Capture Integration} 
& Integrate lightweight evidence capture and reuse to reduce documentation burden. 
Stakeholders need minimal manual input, reuse of existing artifacts, and audit-ready summaries for external review. 
& Evidence nodes or parameters; auto-ingestion pipelines; exportable evidence reports. \\

\textbf{Incentives \& Value Signaling} 
& Include readiness indicators and shareable artifacts aligned with what hubs, funders,  regulators etc. expect. 
Stakeholders need credible indicators that communicate progress and justify continued support or investment. 
& Milestone badges; readiness summaries; hub-facing reports. \\

\textbf{Iterative Co-Design} 
& Use task-based evaluation and iterative co-design to refine dashboard clarity and fit. 

& Prototype dashboards; task cards; annotated mock-ups; feedback logs. \\

\textbf{Validation Across Contexts} 
& Validate dashboards with diverse stakeholders to ensure robustness across interpretive communities. 

& Multi-stakeholder validation sessions. \\

\textbf{Sustainability \& Reuse} 
& Treat the knowledge and dashboard components as reusable infrastructure supporting long-term governance. 
Stakeholders require stable, updateable tools that maintain continuity across teams and funding cycles. 
& Versioned KG; template library. \\
\bottomrule
\end{tabular}
\caption{Design and validation considerations for governance dashboards}
\label{table}
\end{table*}

\subsection{ Seeing humans behind the tasks}
Across the design studies, the case-study, and the startup survey, RAI work experiences were consistently linked to emotional load. Founders and project leads described governance obligations as anxiety-inducing because expectations were unclear, consequences felt high-stakes, and timing tradeoffs were acute. The cohort emotion landscape captures the co-presence of ambition and overload; in this context, dashboards function as a supporting tool aiming to reduce uncertainty, redistributing it across visible, staged tasks. This also can draw implications for visual design patterns aimed at reducing stress and reinforcing sense of progression.   

Inclusive visualization design matters for participation in these decisions: age, skill, and accessibility constraints can determine whether clinicians and non-technical roles can meaningfully engage with governance evidence. This human experience dimension connects to diversity in innovation. When governance overhead is opaque and costly, it disproportionately disadvantages teams with fewer resources, first-time founders, or otherwise disadvantaged groups. Shared infrastructure therefore can lower barriers for teams pursuing clinically and socially valuable problems that may be less commercially straightforward, and can support plurality in innovation by making evidence practices more attainable for resource-constrained teams.

\subsection{Policy and governance implications}

Our evidence supports a policy recommendation: early-stage governance capacity is a bottleneck, and policymakers and ecosystem actors can reduce it by promoting interpretable, reusable RAI infrastructure that supports decision-making throughout innovation, rather than relying on static guidance and after-the-fact review. In our research, governance work was project-specific by design; however, processes, assets, and design patterns repeatedly recurred across projects, enabling reuse without enforcing uniformity. The survey cohort showed strong openness to shared tooling and readiness assessment when organized through a hub. This points to a practical recommendation for public funders and regulators: align incentives by requesting staged, TRL-appropriate ongoing milestone reporting, and by providing shared governance infrastructure, shifting oversight from episodic enforcement to continuous capacity-building. Governance dashboards can be piloted as intermediary artifacts in translational programs and regulatory sandboxes, where teams demonstrate evolving evidence readiness over time, aligned to maturity and context, while using the same artifacts internally to coordinate accountability and evidence production. The primary implementation constraint here is sustained coordination; within these bounds, dashboard-based infrastructure provides a concrete mechanism for operationalising compliance by design in health AI.


\section{Conclusion}

This paper examined how governance dashboards can be designed to support Responsible AI practice in early-stage HealthTech innovation, where regulatory expectations often exceed organizational capacity. Through longitudinal design study insights, cohort-based survey, and a case study, we explore how Responsible AI dashboard design can be conducted and what is needed to make them practical and sustainable. More broadly, this work lays groundwork for further research in visual governance infrastructure in enabling Responsible AI adoption within innovation hub community, and calls for future research that examines how such tools interact with institutional incentives and regulatory practice. Our future work focuses on the implementation and validation of the design practices within innovation hubs that could contribute to better formulation of collaborative design approaches and to building value-led, diverse AI innovation capacity at national and regional levels.

\textbf{Limitations.} This study is situated within the UK and EU regulatory context for healthcare, and its findings should be interpreted accordingly. We invite future work to explore generalisation to other domains with early-stage AI innovation, such as education or finance. There are risks that, without continual iteration, a governance dashboard could become outdated or could inadvertently encourage a minimal compliance mindset (if teams focus only on the items shown and ignore issues outside its scope), and hence we focus on the design study as a process, rather than offering specific dashboards or solutions. Responsible AI infrastructure must itself be maintained responsibly: staying agile and rooted in the realities of real-world operation through user feedback loops, ongoing feedback from domain experts, and integration of new regulatory or ethical insights as they emerge.

\section{Positionality Statement}
The authors work at the intersection of academia and industry and collaborate closely with health-technology innovators, accelerator organizations, and governance practitioners. This positioning provides direct exposure to the operational constraints and decision-making realities of early-stage HealthTech teams, enabling iterative feedback throughout the design and evaluation of governance dashboards. Our team spans backgrounds in human–computer interaction, visualization, clinical innovation, and Responsible AI regulation, which affords multiple disciplinary viewpoints on governance challenges. Although these partnerships strengthen ecological validity, we recognize that our proximity to industry stakeholders may shape how we interpret feasibility, workflow fit, and organizational priorities. We address this by engaging diverse participants, including startups, innovation hubs, research organizations, and by foregrounding plural interpretations and context-specific constraints in our methodological choices.

\section{Generative AI Usage Statement}
Authors acknowledge the use of Otter.ai (https://otter.ai/) for interview recording transcription and ChatGPT v.5.2 (https://chat.openai.com/) for grammar and style editing and to prototype visual layouts that subsequently were manually edited. The authors take full responsibility for the accuracy and originality of the submission.

\bibliographystyle{ACM-Reference-Format}
\bibliography{references}


\begin{thebibliography}{75}


\ifx \showCODEN    \undefined \def \showCODEN     #1{\unskip}     \fi
\ifx \showISBNx    \undefined \def \showISBNx     #1{\unskip}     \fi
\ifx \showISBNxiii \undefined \def \showISBNxiii  #1{\unskip}     \fi
\ifx \showISSN     \undefined \def \showISSN      #1{\unskip}     \fi
\ifx \showLCCN     \undefined \def \showLCCN      #1{\unskip}     \fi
\ifx \shownote     \undefined \def \shownote      #1{#1}          \fi
\ifx \showarticletitle \undefined \def \showarticletitle #1{#1}   \fi
\ifx \showURL      \undefined \def \showURL       {\relax}        \fi
\providecommand\bibfield[2]{#2}
\providecommand\bibinfo[2]{#2}
\providecommand\natexlab[1]{#1}
\providecommand\showeprint[2][]{arXiv:#2}

\bibitem[Akbaba et~al\mbox{.}(2023)]%
        {Akbaba2023}
\bibfield{author}{\bibinfo{person}{Derya Akbaba}, \bibinfo{person}{Devin Lange}, \bibinfo{person}{Michael Correll}, \bibinfo{person}{Alexander Lex}, {and} \bibinfo{person}{Miriah Meyer}.} \bibinfo{year}{2023}\natexlab{}.
\newblock \showarticletitle{{Troubling Collaboration: Matters of Care for Visualization Design Study}}. In \bibinfo{booktitle}{\emph{Conference on Human Factors in Computing Systems - Proceedings}}. \bibinfo{publisher}{Association for Computing Machinery}.
\newblock
\showISBNx{9781450394215}
\href{https://doi.org/10.1145/3544548.3581168}{doi:\nolinkurl{10.1145/3544548.3581168}}


\bibitem[Alanoca et~al\mbox{.}(2025)]%
        {Alanoca2025}
\bibfield{author}{\bibinfo{person}{Sacha Alanoca}, \bibinfo{person}{Shira Gur-Arieh}, \bibinfo{person}{Tom Zick}, {and} \bibinfo{person}{Kevin Klyman}.} \bibinfo{year}{2025}\natexlab{}.
\newblock \showarticletitle{{Comparing Apples to Oranges: A Taxonomy for Navigating the Global Landscape of AI Regulation}}. In \bibinfo{booktitle}{\emph{ACMF AccT 2025 - Proceedings of the 2025 ACM Conference on Fairness, Accountability,and Transparency}}. \bibinfo{publisher}{Association for Computing Machinery, Inc}, \bibinfo{pages}{914--937}.
\newblock
\showISBNx{9798400714825}
\href{https://doi.org/10.1145/3715275.3732059}{doi:\nolinkurl{10.1145/3715275.3732059}}


\bibitem[Alpherts et~al\mbox{.}(2024)]%
        {Alpherts2024}
\bibfield{author}{\bibinfo{person}{Tim Alpherts}, \bibinfo{person}{Sennay Ghebreab}, \bibinfo{person}{Yen~Chia Hsu}, {and} \bibinfo{person}{Nanne Van~Noord}.} \bibinfo{year}{2024}\natexlab{}.
\newblock \showarticletitle{{Perceptive Visual Urban Analytics is Not (Yet) Suitable for Municipalities}}. In \bibinfo{booktitle}{\emph{2024 ACM Conference on Fairness, Accountability, and Transparency, FAccT 2024}}. \bibinfo{publisher}{Association for Computing Machinery, Inc}, \bibinfo{pages}{1341--1354}.
\newblock
\showISBNx{9798400704505}
\href{https://doi.org/10.1145/3630106.3658976}{doi:\nolinkurl{10.1145/3630106.3658976}}


\bibitem[Amazon Web~Services(2022)]%
        {AWSAI}
\bibfield{author}{\bibinfo{person}{Inc. Amazon Web~Services}.} \bibinfo{year}{2022}\natexlab{}.
\newblock \bibinfo{booktitle}{\emph{{Responsible Use of Machine Learning}}}.
\newblock \bibinfo{type}{{T}echnical {R}eport}. \bibinfo{institution}{Amazon Web Services}.
\newblock
\urldef\tempurl%
\url{https://aws.amazon.com/ai/responsible-ai/}
\showURL{%
\tempurl}


\bibitem[Antweiler and Fuchs(2022)]%
        {Antweiler2022}
\bibfield{author}{\bibinfo{person}{Dario Antweiler} {and} \bibinfo{person}{Georg Fuchs}.} \bibinfo{year}{2022}\natexlab{}.
\newblock \showarticletitle{{Visualizing Rule-based Classifiers for Clinical Risk Prognosis}}. In \bibinfo{booktitle}{\emph{Proceedings - 2022 IEEE Visualization Conference - Short Papers, VIS 2022}}. \bibinfo{publisher}{Institute of Electrical and Electronics Engineers Inc.}, \bibinfo{pages}{55--59}.
\newblock
\showISBNx{9781665488129}
\href{https://doi.org/10.1109/VIS54862.2022.00020}{doi:\nolinkurl{10.1109/VIS54862.2022.00020}}


\bibitem[{Atlassian Pty Ltd.}(2025)]%
        {Jira2025}
\bibfield{author}{\bibinfo{person}{{Atlassian Pty Ltd.}}} \bibinfo{year}{2025}\natexlab{}.
\newblock \bibinfo{title}{{Jira Software}}.
\newblock
\urldef\tempurl%
\url{https://www.atlassian.com/software/jira}
\showURL{%
\tempurl}


\bibitem[Baquero et~al\mbox{.}(2020)]%
        {Baquero2020}
\bibfield{author}{\bibinfo{person}{Juan~Aristi Baquero}, \bibinfo{person}{Roger Burkhardt}, \bibinfo{person}{Arvind Govindarajan}, {and} \bibinfo{person}{Thomas Wallace}.} \bibinfo{year}{2020}\natexlab{}.
\newblock \bibinfo{booktitle}{\emph{{Derisking AI by design: How to build risk management into AI development}}}.
\newblock \bibinfo{type}{{T}echnical {R}eport}.
\newblock


\bibitem[{Brian Kalis and Cassandra Lea and Andrew Truscott}(2025)]%
        {Accenture}
\bibfield{author}{\bibinfo{person}{{Brian Kalis and Cassandra Lea and Andrew Truscott}}.} \bibinfo{year}{2025}\natexlab{}.
\newblock \bibinfo{title}{{Accenture Technology Vision 2025 and Healthcare}}.
\newblock
\urldef\tempurl%
\url{https://www.accenture.com/us-en/blogs/health/accenture-technology-trends-2025-healthcare}
\showURL{%
\tempurl}


\bibitem[By et~al\mbox{.}(2019)]%
        {McKinsey}
\bibfield{author}{\bibinfo{person}{Article By}, \bibinfo{person}{Jake Silberg}, {and} \bibinfo{person}{James Manyika}.} \bibinfo{year}{2019}\natexlab{}.
\newblock \bibinfo{booktitle}{\emph{{Notes from the AI frontier: Tackling bias in AI (and in humans)}}}.
\newblock \bibinfo{type}{{T}echnical {R}eport}.
\newblock


\bibitem[Chen et~al\mbox{.}(2023)]%
        {Chen2023}
\bibfield{author}{\bibinfo{person}{You Chen}, \bibinfo{person}{Ellen~Wright Clayton}, \bibinfo{person}{Laurie~Lovett Novak}, \bibinfo{person}{Shilo Anders}, {and} \bibinfo{person}{Bradley Malin}.} \bibinfo{year}{2023}\natexlab{}.
\newblock \showarticletitle{{Human-Centered Design to Address Biases in Artificial Intelligence}}.
\newblock \bibinfo{journal}{\emph{Journal of Medical Internet Research}}  \bibinfo{volume}{25} (\bibinfo{date}{3} \bibinfo{year}{2023}), \bibinfo{pages}{e43251}.
\newblock
\showISSN{1438-8871}
\href{https://doi.org/10.2196/43251}{doi:\nolinkurl{10.2196/43251}}


\bibitem[Chomutare et~al\mbox{.}(2022)]%
        {CTS2022}
\bibfield{author}{\bibinfo{person}{Taridzo Chomutare}, \bibinfo{person}{Miguel Tejedor}, \bibinfo{person}{Therese~Olsen Svenning}, \bibinfo{person}{Luis Marco-Ruiz}, \bibinfo{person}{Maryam Tayefi}, \bibinfo{person}{Karianne Lind}, \bibinfo{person}{Fred Godtliebsen}, \bibinfo{person}{Anne Moen}, \bibinfo{person}{Leila Ismail}, \bibinfo{person}{Alexandra Makhlysheva}, {and} \bibinfo{person}{Phuong~Dinh Ngo}.} \bibinfo{year}{2022}\natexlab{}.
\newblock \showarticletitle{{Artificial Intelligence Implementation in Healthcare: A Theory-Based Scoping Review of Barriers and Facilitators}}.
\newblock \bibinfo{journal}{\emph{International Journal of Environmental Research and Public Health}} \bibinfo{volume}{19}, \bibinfo{number}{23} (\bibinfo{year}{2022}).
\newblock
\showISSN{1660-4601}
\href{https://doi.org/10.3390/ijerph192316359}{doi:\nolinkurl{10.3390/ijerph192316359}}


\bibitem[Cibulski et~al\mbox{.}(2022)]%
        {Cibulski2022}
\bibfield{author}{\bibinfo{person}{Lena Cibulski}, \bibinfo{person}{Evanthia Dimara}, \bibinfo{person}{Setia Hermawati}, {and} \bibinfo{person}{Jorn Kohlhammer}.} \bibinfo{year}{2022}\natexlab{}.
\newblock \showarticletitle{{Supporting Domain Characterization in Visualization Design Studies With the Critical Decision Method}}. In \bibinfo{booktitle}{\emph{Proceedings - 2022 IEEE 4th Workshop on Visualization Guidelines in Research, Design, and Education, VisGuides 2022}}. \bibinfo{publisher}{Institute of Electrical and Electronics Engineers Inc.}, \bibinfo{pages}{8--15}.
\newblock
\showISBNx{9798350397123}
\href{https://doi.org/10.1109/VisGuides57787.2022.00007}{doi:\nolinkurl{10.1109/VisGuides57787.2022.00007}}


\bibitem[De et~al\mbox{.}(2025)]%
        {De2025}
\bibfield{author}{\bibinfo{person}{Ankolika De}, \bibinfo{person}{Shaheen Kanthawala}, {and} \bibinfo{person}{Jessica Maddox}.} \bibinfo{year}{2025}\natexlab{}.
\newblock \showarticletitle{{Who Gets Heard? Calling Out the "hard-To-Reach" Myth for Non-WEIRD Populations' Recruitment and Involvement in Research}}. In \bibinfo{booktitle}{\emph{ACMF AccT 2025 - Proceedings of the 2025 ACM Conference on Fairness, Accountability,and Transparency}}. \bibinfo{publisher}{Association for Computing Machinery, Inc}, \bibinfo{pages}{855--867}.
\newblock
\showISBNx{9798400714825}
\href{https://doi.org/10.1145/3715275.3732055}{doi:\nolinkurl{10.1145/3715275.3732055}}


\bibitem[De~Troya et~al\mbox{.}(2025)]%
        {deTroya2025}
\bibfield{author}{\bibinfo{person}{Íñigo De~Troya}, \bibinfo{person}{Jacqueline Kernahan}, \bibinfo{person}{Neelke Doorn}, \bibinfo{person}{Virginia Dignum}, {and} \bibinfo{person}{Roel Dobbe}.} \bibinfo{year}{2025}\natexlab{}.
\newblock \showarticletitle{{Misabstraction in Sociotechnical Systems}}. In \bibinfo{booktitle}{\emph{ACMF AccT 2025 - Proceedings of the 2025 ACM Conference on Fairness, Accountability,and Transparency}}. \bibinfo{publisher}{Association for Computing Machinery, Inc}, \bibinfo{pages}{1829--1842}.
\newblock
\showISBNx{9798400714825}
\href{https://doi.org/10.1145/3715275.3732122}{doi:\nolinkurl{10.1145/3715275.3732122}}


\bibitem[{European Commission}(2017)]%
        {EuropeanCommission2017}
\bibfield{author}{\bibinfo{person}{{European Commission}}.} \bibinfo{year}{2017}\natexlab{}.
\newblock \bibinfo{booktitle}{\emph{{Technology Readiness Level: Guidance Principles for Renewable Energy technologies Final Report}}}.
\newblock \bibinfo{type}{{T}echnical {R}eport}. \bibinfo{institution}{Directorate-General for Research and Innovation, Publications Office of the European Union}, \bibinfo{address}{Luxembourg}.
\newblock
\href{https://doi.org/10.2777/577767}{doi:\nolinkurl{10.2777/577767}}


\bibitem[{European Union}(2024)]%
        {EUAIAct}
\bibfield{author}{\bibinfo{person}{{European Union}}.} \bibinfo{year}{2024}\natexlab{}.
\newblock \bibinfo{booktitle}{\emph{{Regulation (EU) 2024/1689 of the European Parliament and of the Council laying down harmonised rules on artificial intelligence (Artificial Intelligence Act)}}}.
\newblock \bibinfo{type}{{T}echnical {R}eport}. \bibinfo{institution}{Official Journal of the European Union}.
\newblock


\bibitem[Gerke et~al\mbox{.}(2020)]%
        {Gerke2020}
\bibfield{author}{\bibinfo{person}{Sara Gerke}, \bibinfo{person}{Timo Minssen}, {and} \bibinfo{person}{Glenn Cohen}.} \bibinfo{year}{2020}\natexlab{}.
\newblock \showarticletitle{{Ethical and legal challenges of artificial intelligence-driven healthcare}}.
\newblock In \bibinfo{booktitle}{\emph{Artificial Intelligence in Healthcare}}. \bibinfo{publisher}{Elsevier}, \bibinfo{pages}{295--336}.
\newblock
\showISBNx{9780128184387}
\href{https://doi.org/10.1016/B978-0-12-818438-7.00012-5}{doi:\nolinkurl{10.1016/B978-0-12-818438-7.00012-5}}


\bibitem[Greenes et~al\mbox{.}(2018)]%
        {Greenes2018}
\bibfield{author}{\bibinfo{person}{Robert~A. Greenes}, \bibinfo{person}{David~W. Bates}, \bibinfo{person}{Kensaku Kawamoto}, \bibinfo{person}{Blackford Middleton}, \bibinfo{person}{Jerome Osheroff}, {and} \bibinfo{person}{Yuval Shahar}.} \bibinfo{year}{2018}\natexlab{}.
\newblock \showarticletitle{{Clinical decision support models and frameworks: Seeking to address research issues underlying implementation successes and failures}}.
\newblock \bibinfo{journal}{\emph{Journal of Biomedical Informatics}}  \bibinfo{volume}{78} (\bibinfo{date}{2} \bibinfo{year}{2018}), \bibinfo{pages}{134--143}.
\newblock
\showISSN{15320464}
\href{https://doi.org/10.1016/j.jbi.2017.12.005}{doi:\nolinkurl{10.1016/j.jbi.2017.12.005}}


\bibitem[Hall et~al\mbox{.}(2020)]%
        {Hall2020}
\bibfield{author}{\bibinfo{person}{Kyle~Wm Hall}, \bibinfo{person}{Adam~J. Bradley}, \bibinfo{person}{Uta Hinrichs}, \bibinfo{person}{Samuel Huron}, \bibinfo{person}{Jo Wood}, \bibinfo{person}{Christopher Collins}, {and} \bibinfo{person}{Sheelagh Carpendale}.} \bibinfo{year}{2020}\natexlab{}.
\newblock \showarticletitle{{Design by Immersion: A Transdisciplinary Approach to Problem-Driven Visualizations}}.
\newblock \bibinfo{journal}{\emph{IEEE Transactions on Visualization and Computer Graphics}} \bibinfo{volume}{26}, \bibinfo{number}{1} (\bibinfo{date}{1} \bibinfo{year}{2020}), \bibinfo{pages}{109--118}.
\newblock
\showISSN{19410506}
\href{https://doi.org/10.1109/TVCG.2019.2934790}{doi:\nolinkurl{10.1109/TVCG.2019.2934790}}


\bibitem[Hollanek et~al\mbox{.}(2025)]%
        {Hollanek2025}
\bibfield{author}{\bibinfo{person}{Tomasz Hollanek}, \bibinfo{person}{Yulu Pi}, \bibinfo{person}{Cosimo Fiorini}, \bibinfo{person}{Virginia Vignali}, \bibinfo{person}{Dorian Peters}, {and} \bibinfo{person}{Eleanor Drage}.} \bibinfo{year}{2025}\natexlab{}.
\newblock \showarticletitle{{A Toolkit for Compliance, a Toolkit for Justice: Drawing on Cross-sectoral Expertise to Develop a Pro-justice EU AI Act Toolkit}}. In \bibinfo{booktitle}{\emph{ACMF AccT 2025 - Proceedings of the 2025 ACM Conference on Fairness, Accountability,and Transparency}}. \bibinfo{publisher}{Association for Computing Machinery, Inc}, \bibinfo{pages}{1184--1194}.
\newblock
\showISBNx{9798400714825}
\href{https://doi.org/10.1145/3715275.3732078}{doi:\nolinkurl{10.1145/3715275.3732078}}


\bibitem[Hopkins et~al\mbox{.}(2025)]%
        {Hopkins2025}
\bibfield{author}{\bibinfo{person}{Aspen Hopkins}, \bibinfo{person}{Isabella Struckman}, \bibinfo{person}{Kevin Klyman}, {and} \bibinfo{person}{Susan~S. Silbey}.} \bibinfo{year}{2025}\natexlab{}.
\newblock \showarticletitle{{Recourse, Repair, Reparation, {\&} Prevention: A Stakeholder Analysis of AI Supply Chains}}. In \bibinfo{booktitle}{\emph{ACMF AccT 2025 - Proceedings of the 2025 ACM Conference on Fairness, Accountability,and Transparency}}. \bibinfo{publisher}{Association for Computing Machinery, Inc}, \bibinfo{pages}{209--227}.
\newblock
\showISBNx{9798400714825}
\href{https://doi.org/10.1145/3715275.3732017}{doi:\nolinkurl{10.1145/3715275.3732017}}


\bibitem[{IEEE}(2024)]%
        {IEEE-standard2024}
\bibfield{author}{\bibinfo{person}{{IEEE}}.} \bibinfo{year}{2024}\natexlab{}.
\newblock \bibinfo{title}{{IEEE Standard for Ethical Considerations in Emulated Empathy in Autonomous and Intelligent Systems}}.
\newblock
\showISBNx{979-8-8557-0849-3}
\href{https://doi.org/10.1109/IEEESTD.2024.10576666}{doi:\nolinkurl{10.1109/IEEESTD.2024.10576666}}


\bibitem[Isenberg et~al\mbox{.}(2011)]%
        {Isenberg2011}
\bibfield{author}{\bibinfo{person}{Petra Isenberg}, \bibinfo{person}{Niklas Elmqvist}, \bibinfo{person}{Jean Scholtz}, \bibinfo{person}{Daniel Cernea}, \bibinfo{person}{Kwan-Liu Ma}, {and} \bibinfo{person}{Hans Hagen}.} \bibinfo{year}{2011}\natexlab{}.
\newblock \showarticletitle{{Collaborative visualization: Definition, challenges, and research agenda}}.
\newblock \bibinfo{journal}{\emph{Information Visualization}} \bibinfo{volume}{10}, \bibinfo{number}{4} (\bibinfo{year}{2011}), \bibinfo{pages}{310--326}.
\newblock
\href{https://doi.org/10.1177/1473871611412817}{doi:\nolinkurl{10.1177/1473871611412817}}


\bibitem[J{\"{a}}nicke et~al\mbox{.}(2020)]%
        {Janicke2020}
\bibfield{author}{\bibinfo{person}{S. J{\"{a}}nicke}, \bibinfo{person}{P. Kaur}, \bibinfo{person}{P. K{\'{u}}zmicki}, {and} \bibinfo{person}{J. Schmidt}.} \bibinfo{year}{2020}\natexlab{}.
\newblock \showarticletitle{{Participatory Visualization Design as an Approach to Minimize the Gap between Research and Application}}. In \bibinfo{booktitle}{\emph{VisGap 2020 - Gap between Visualization Research and Visualization Software}}. \bibinfo{publisher}{The Eurographics Association}, \bibinfo{pages}{35--42}.
\newblock
\showISBNx{9783038681250}
\href{https://doi.org/10.2312/visgap.20201108}{doi:\nolinkurl{10.2312/visgap.20201108}}


\bibitem[Johnson et~al\mbox{.}(2025)]%
        {Johnson2025}
\bibfield{author}{\bibinfo{person}{Nari Johnson}, \bibinfo{person}{Elise Silva}, \bibinfo{person}{Harrison Leon}, \bibinfo{person}{Motahhare Eslami}, \bibinfo{person}{Beth Schwanke}, \bibinfo{person}{Ravit Dotan}, {and} \bibinfo{person}{Hoda Heidari}.} \bibinfo{year}{2025}\natexlab{}.
\newblock \showarticletitle{{Legacy Procurement Practices Shape How U.S. Cities Govern AI: Understanding Government Employees' Practices, Challenges, and Needs}}. In \bibinfo{booktitle}{\emph{ACMF AccT 2025 - Proceedings of the 2025 ACM Conference on Fairness, Accountability,and Transparency}}. \bibinfo{publisher}{Association for Computing Machinery, Inc}, \bibinfo{pages}{772--789}.
\newblock
\showISBNx{9798400714825}
\href{https://doi.org/10.1145/3715275.3732049}{doi:\nolinkurl{10.1145/3715275.3732049}}


\bibitem[Kerzner et~al\mbox{.}(2019)]%
        {Kerzner2020}
\bibfield{author}{\bibinfo{person}{Ethan Kerzner}, \bibinfo{person}{Sarah Goodwin}, \bibinfo{person}{Jason Dykes}, \bibinfo{person}{Sara Jones}, {and} \bibinfo{person}{Miriah Meyer}.} \bibinfo{year}{2019}\natexlab{}.
\newblock \showarticletitle{{A Framework for Creative Visualization-Opportunities Workshops}}.
\newblock \bibinfo{journal}{\emph{IEEE Transactions on Visualization and Computer Graphics}} \bibinfo{volume}{25}, \bibinfo{number}{1} (\bibinfo{date}{1} \bibinfo{year}{2019}), \bibinfo{pages}{748--758}.
\newblock
\showISSN{19410506}
\href{https://doi.org/10.1109/TVCG.2018.2865241}{doi:\nolinkurl{10.1109/TVCG.2018.2865241}}


\bibitem[Kim et~al\mbox{.}(2021)]%
        {Kim2021}
\bibfield{author}{\bibinfo{person}{N~W Kim}, \bibinfo{person}{S~C Joyner}, \bibinfo{person}{A Riegelhuth}, {and} \bibinfo{person}{Y Kim}.} \bibinfo{year}{2021}\natexlab{}.
\newblock \showarticletitle{{Accessible Visualization: Design Space, Opportunities, and Challenges}}.
\newblock \bibinfo{journal}{\emph{Computer Graphics Forum}} \bibinfo{volume}{40}, \bibinfo{number}{3} (\bibinfo{year}{2021}), \bibinfo{pages}{173--188}.
\newblock
\href{https://doi.org/10.1111/cgf.14298}{doi:\nolinkurl{10.1111/cgf.14298}}


\bibitem[Kinney(2025)]%
        {Kinney2025}
\bibfield{author}{\bibinfo{person}{David Kinney}.} \bibinfo{year}{2025}\natexlab{}.
\newblock \showarticletitle{{Aggregating Concepts of Fairness and Accuracy in Prediction Algorithms}}. In \bibinfo{booktitle}{\emph{ACMF AccT 2025 - Proceedings of the 2025 ACM Conference on Fairness, Accountability,and Transparency}}. \bibinfo{publisher}{Association for Computing Machinery, Inc}, \bibinfo{pages}{464--472}.
\newblock
\showISBNx{9798400714825}
\href{https://doi.org/10.1145/3715275.3732031}{doi:\nolinkurl{10.1145/3715275.3732031}}


\bibitem[Liang and Wang(2025)]%
        {Liang2025}
\bibfield{author}{\bibinfo{person}{Jia~Wei Liang} {and} \bibinfo{person}{Hao~Chuan Wang}.} \bibinfo{year}{2025}\natexlab{}.
\newblock \showarticletitle{{Is It Fair Enough? Supporting Equitable Group Work Assignment with Work Division Dashboard}}. In \bibinfo{booktitle}{\emph{ACMF AccT 2025 - Proceedings of the 2025 ACM Conference on Fairness, Accountability,and Transparency}}. \bibinfo{publisher}{Association for Computing Machinery, Inc}, \bibinfo{pages}{2480--2490}.
\newblock
\showISBNx{9798400714825}
\href{https://doi.org/10.1145/3715275.3732163}{doi:\nolinkurl{10.1145/3715275.3732163}}


\bibitem[Lovelace~Institute and on({[n.\,d.]})]%
        {LovelaceInstitute}
\bibfield{author}{\bibinfo{person}{Ada Lovelace~Institute} {and} \bibinfo{person}{Partnership~AI on}.} \bibinfo{year}{[n.\,d.]}\natexlab{}.
\newblock \bibinfo{booktitle}{\emph{{A CULTURE OF ETHICAL AI: REPORT}}}.
\newblock \bibinfo{type}{{T}echnical {R}eport}.
\newblock


\bibitem[Lucchesi et~al\mbox{.}(2022)]%
        {Lucchesi2022}
\bibfield{author}{\bibinfo{person}{Lydia~R. Lucchesi}, \bibinfo{person}{Petra~M. Kuhnert}, \bibinfo{person}{Jenny~L. Davis}, {and} \bibinfo{person}{Lexing Xie}.} \bibinfo{year}{2022}\natexlab{}.
\newblock \showarticletitle{{Smallset Timelines: A Visual Representation of Data Preprocessing Decisions}}. In \bibinfo{booktitle}{\emph{ACM International Conference Proceeding Series}}. \bibinfo{publisher}{Association for Computing Machinery}, \bibinfo{pages}{1136--1153}.
\newblock
\showISBNx{9781450393522}
\href{https://doi.org/10.1145/3531146.3533175}{doi:\nolinkurl{10.1145/3531146.3533175}}


\bibitem[Lundgard et~al\mbox{.}(2019)]%
        {Lundgard2019}
\bibfield{author}{\bibinfo{person}{Alan Lundgard}, \bibinfo{person}{Crystal Lee}, {and} \bibinfo{person}{Arvind Satyanarayan}.} \bibinfo{year}{2019}\natexlab{}.
\newblock \showarticletitle{{Sociotechnical Considerations for Accessible Visualization Design}}. In \bibinfo{booktitle}{\emph{2019 IEEE Visualization Conference (VIS)}}. \bibinfo{pages}{16--20}.
\newblock
\href{https://doi.org/10.1109/VISUAL.2019.8933762}{doi:\nolinkurl{10.1109/VISUAL.2019.8933762}}


\bibitem[Marai(2018)]%
        {Marai2018}
\bibfield{author}{\bibinfo{person}{G.~Elisabeta Marai}.} \bibinfo{year}{2018}\natexlab{}.
\newblock \showarticletitle{{Activity-Centered Domain Characterization for Problem-Driven Scientific Visualization}}.
\newblock \bibinfo{journal}{\emph{IEEE Transactions on Visualization and Computer Graphics}} \bibinfo{volume}{24}, \bibinfo{number}{1} (\bibinfo{date}{1} \bibinfo{year}{2018}), \bibinfo{pages}{913--922}.
\newblock
\showISSN{10772626}
\href{https://doi.org/10.1109/TVCG.2017.2744459}{doi:\nolinkurl{10.1109/TVCG.2017.2744459}}


\bibitem[Mckenna et~al\mbox{.}(2014)]%
        {MMA2014}
\bibfield{author}{\bibinfo{person}{Sean Mckenna}, \bibinfo{person}{David Mazur}, \bibinfo{person}{James Agutter}, {and} \bibinfo{person}{Michael Meyer}.} \bibinfo{year}{2014}\natexlab{}.
\newblock \showarticletitle{{Design Activity Framework for Visualization Design}}.
\newblock \bibinfo{journal}{\emph{IEEE Transactions on Visualization and Computer Graphics}}  \bibinfo{volume}{20} (\bibinfo{date}{1} \bibinfo{year}{2014}), \bibinfo{pages}{2191--2200}.
\newblock
\href{https://doi.org/10.1109/TVCG.2014.2346331}{doi:\nolinkurl{10.1109/TVCG.2014.2346331}}


\bibitem[Moore et~al\mbox{.}(2025)]%
        {Moore2025}
\bibfield{author}{\bibinfo{person}{Jared Moore}, \bibinfo{person}{Declan Grabb}, \bibinfo{person}{William Agnew}, \bibinfo{person}{Kevin Klyman}, \bibinfo{person}{Stevie Chancellor}, \bibinfo{person}{Desmond~C. Ong}, {and} \bibinfo{person}{Nick Haber}.} \bibinfo{year}{2025}\natexlab{}.
\newblock \showarticletitle{{Expressing stigma and inappropriate responses prevents LLMs from safely replacing mental health providers.}}. In \bibinfo{booktitle}{\emph{ACMF AccT 2025 - Proceedings of the 2025 ACM Conference on Fairness, Accountability,and Transparency}}. \bibinfo{publisher}{Association for Computing Machinery, Inc}, \bibinfo{pages}{599--627}.
\newblock
\showISBNx{9798400714825}
\href{https://doi.org/10.1145/3715275.3732039}{doi:\nolinkurl{10.1145/3715275.3732039}}


\bibitem[Morley et~al\mbox{.}(2020)]%
        {Morley2020}
\bibfield{author}{\bibinfo{person}{Jessica Morley}, \bibinfo{person}{Caio~C.V. Machado}, \bibinfo{person}{Christopher Burr}, \bibinfo{person}{Josh Cowls}, \bibinfo{person}{Indra Joshi}, \bibinfo{person}{Mariarosaria Taddeo}, {and} \bibinfo{person}{Luciano Floridi}.} \bibinfo{year}{2020}\natexlab{}.
\newblock \bibinfo{title}{{The ethics of AI in health care: A mapping review}}.
\newblock
\showISSN{18735347}
\href{https://doi.org/10.1016/j.socscimed.2020.113172}{doi:\nolinkurl{10.1016/j.socscimed.2020.113172}}


\bibitem[Mueller et~al\mbox{.}(2020)]%
        {Mueller2020}
\bibfield{author}{\bibinfo{person}{Shane~T Mueller}, \bibinfo{person}{Elizabeth~S Veinott}, \bibinfo{person}{Robert~R Hoffman}, \bibinfo{person}{Gary Klein}, \bibinfo{person}{Lamia Alam}, \bibinfo{person}{Tauseef Mamun}, {and} \bibinfo{person}{William~J Clancey}.} \bibinfo{year}{2020}\natexlab{}.
\newblock \bibinfo{booktitle}{\emph{{Principles of Explanation in Human-AI Systems}}}.
\newblock \bibinfo{type}{{T}echnical {R}eport}.
\newblock
\urldef\tempurl%
\url{www.aaai.org}
\showURL{%
\tempurl}


\bibitem[Munzner(2009)]%
        {MUN2009}
\bibfield{author}{\bibinfo{person}{Tamara Munzner}.} \bibinfo{year}{2009}\natexlab{}.
\newblock \showarticletitle{{A Nested Model for Visualization Design and Validation}}.
\newblock \bibinfo{journal}{\emph{IEEE Transactions on Visualization and Computer Graphics}}  \bibinfo{volume}{15} (\bibinfo{year}{2009}), \bibinfo{pages}{921--928}.
\newblock
\href{https://doi.org/10.1109/TVCG.2009.111}{doi:\nolinkurl{10.1109/TVCG.2009.111}}


\bibitem[Munzner(2014)]%
        {munzner2014visualization}
\bibfield{author}{\bibinfo{person}{Tamara Munzner}.} \bibinfo{year}{2014}\natexlab{}.
\newblock \bibinfo{booktitle}{\emph{{Visualization Analysis and Design}}}.
\newblock \bibinfo{publisher}{A K Peters/CRC Press}.
\newblock
\showISBNx{9780429088902}
\href{https://doi.org/10.1201/b17511}{doi:\nolinkurl{10.1201/b17511}}


\bibitem[Neo4j(2025)]%
        {Neo4j2025}
\bibfield{author}{\bibinfo{person}{Inc. Neo4j}.} \bibinfo{year}{2025}\natexlab{}.
\newblock \bibinfo{title}{{Neo4j Graph Data Platform}}.
\newblock
\urldef\tempurl%
\url{https://neo4j.com}
\showURL{%
\tempurl}


\bibitem[Neri et~al\mbox{.}(2025)]%
        {Neri2025}
\bibfield{author}{\bibinfo{person}{Giulia Neri}, \bibinfo{person}{Shevyn Marshall}, \bibinfo{person}{Harry Kai-Ho Chan}, \bibinfo{person}{Abdallah Yaghi}, \bibinfo{person}{Dash Tabor}, \bibinfo{person}{Rahul Sinha}, {and} \bibinfo{person}{Suvodeep Mazumdar}.} \bibinfo{year}{2025}\natexlab{}.
\newblock \showarticletitle{{Data visualization in AI-assisted decision-making: a systematic review}}.
\newblock \bibinfo{journal}{\emph{Frontiers in Communication}}  \bibinfo{volume}{10} (\bibinfo{date}{8} \bibinfo{year}{2025}).
\newblock
\showISSN{2297-900X}
\href{https://doi.org/10.3389/fcomm.2025.1605655}{doi:\nolinkurl{10.3389/fcomm.2025.1605655}}


\bibitem[{NHS England}(2024)]%
        {England2024}
\bibfield{author}{\bibinfo{person}{{NHS England}}.} \bibinfo{year}{2024}\natexlab{}.
\newblock \bibinfo{title}{{Planning and implementing real-world AI evaluations: lessons from the AI in Health and Care Award}}.
\newblock
\urldef\tempurl%
\url{https://www.england.nhs.uk/publication/planning-and-implementing-real-world-artificial-intelligence-ai-evaluations-lessons-from-the-ai-in-health-and-care-award/}
\showURL{%
\tempurl}


\bibitem[Nolte et~al\mbox{.}(2025)]%
        {nolte2025}
\bibfield{author}{\bibinfo{person}{Henrik Nolte}, \bibinfo{person}{Miriam Rateike}, {and} \bibinfo{person}{Michèle Finck}.} \bibinfo{year}{2025}\natexlab{}.
\newblock \showarticletitle{{Robustness and Cybersecurity in the EU Artificial Intelligence Act}}. In \bibinfo{booktitle}{\emph{Proceedings of the 2025 ACM Conference on Fairness, Accountability, and Transparency}}. \bibinfo{publisher}{ACM}, \bibinfo{address}{New York, NY, USA}, \bibinfo{pages}{283--295}.
\newblock
\showISBNx{9798400714825}
\href{https://doi.org/10.1145/3715275.3732020}{doi:\nolinkurl{10.1145/3715275.3732020}}


\bibitem[Oral et~al\mbox{.}(2023)]%
        {Oral2023}
\bibfield{author}{\bibinfo{person}{Emre Oral}, \bibinfo{person}{Ria Chawla}, \bibinfo{person}{Michel Wijkstra}, \bibinfo{person}{Narges Mahyar}, {and} \bibinfo{person}{Evanthia Dimara}.} \bibinfo{year}{2023}\natexlab{}.
\newblock \showarticletitle{{From Information to Choice: A Critical Inquiry Into Visualization Tools for Decision Making}}.
\newblock \bibinfo{journal}{\emph{IEEE Transactions on Visualization and Computer Graphics}} (\bibinfo{date}{7} \bibinfo{year}{2023}).
\newblock
\href{https://doi.org/10.1109/TVCG.2023.3326593}{doi:\nolinkurl{10.1109/TVCG.2023.3326593}}


\bibitem[Otter.ai(2025)]%
        {Otterai2025}
\bibfield{author}{\bibinfo{person}{Inc. Otter.ai}.} \bibinfo{year}{2025}\natexlab{}.
\newblock \bibinfo{title}{{Otter.ai: Automatic Transcription Service}}.
\newblock
\urldef\tempurl%
\url{https://otter.ai}
\showURL{%
\tempurl}


\bibitem[Panigutti et~al\mbox{.}(2025)]%
        {Panigutti2025}
\bibfield{author}{\bibinfo{person}{Cecilia Panigutti}, \bibinfo{person}{Delia~Fano Yela}, \bibinfo{person}{Lorenzo Porcaro}, \bibinfo{person}{Astrid Bertrand}, {and} \bibinfo{person}{Josep~Soler Garrido}.} \bibinfo{year}{2025}\natexlab{}.
\newblock \showarticletitle{{How to investigate algorithmic-driven risks in online platforms and search engines? A narrative review through the lens of the EU Digital Services Act}}. In \bibinfo{booktitle}{\emph{ACMF AccT 2025 - Proceedings of the 2025 ACM Conference on Fairness, Accountability,and Transparency}}. \bibinfo{publisher}{Association for Computing Machinery, Inc}, \bibinfo{pages}{828--839}.
\newblock
\showISBNx{9798400714825}
\href{https://doi.org/10.1145/3715275.3732052}{doi:\nolinkurl{10.1145/3715275.3732052}}


\bibitem[Petersson et~al\mbox{.}(2022)]%
        {PLN2022}
\bibfield{author}{\bibinfo{person}{Lena Petersson}, \bibinfo{person}{Ingrid Larsson}, \bibinfo{person}{Jens~M. Nygren}, \bibinfo{person}{Per Nilsen}, \bibinfo{person}{Margit Neher}, \bibinfo{person}{Julie~E. Reed}, \bibinfo{person}{Daniel Tyskbo}, {and} \bibinfo{person}{Petra Svedberg}.} \bibinfo{year}{2022}\natexlab{}.
\newblock \showarticletitle{{Challenges to implementing artificial intelligence in healthcare: a qualitative interview study with healthcare leaders in Sweden}}.
\newblock \bibinfo{journal}{\emph{BMC Health Services Research}} \bibinfo{volume}{22}, \bibinfo{number}{1} (\bibinfo{date}{12} \bibinfo{year}{2022}).
\newblock
\showISSN{14726963}
\href{https://doi.org/10.1186/s12913-022-08215-8}{doi:\nolinkurl{10.1186/s12913-022-08215-8}}


\bibitem[Preim and Lawonn(2020)]%
        {Preim2020}
\bibfield{author}{\bibinfo{person}{Bernhard Preim} {and} \bibinfo{person}{Kai Lawonn}.} \bibinfo{year}{2020}\natexlab{}.
\newblock \showarticletitle{{A Survey of Visual Analytics for Public Health}}.
\newblock \bibinfo{journal}{\emph{Computer Graphics Forum}} \bibinfo{volume}{39}, \bibinfo{number}{1} (\bibinfo{date}{2} \bibinfo{year}{2020}), \bibinfo{pages}{543--580}.
\newblock
\showISSN{14678659}
\href{https://doi.org/10.1111/cgf.13891}{doi:\nolinkurl{10.1111/cgf.13891}}


\bibitem[Preim et~al\mbox{.}(2018)]%
        {Preim2018}
\bibfield{author}{\bibinfo{person}{B. Preim}, \bibinfo{person}{T. Ropinski}, {and} \bibinfo{person}{P. Isenberg}.} \bibinfo{year}{2018}\natexlab{}.
\newblock \showarticletitle{{A critical analysis of the evaluation practice in medical visualization}}.
\newblock \bibinfo{journal}{\emph{VCBM 2018 - Eurographics Workshop on Visual Computing for Biology and Medicine}} (\bibinfo{year}{2018}), \bibinfo{pages}{45--56}.
\newblock
\showISBNx{9783038680567}
\href{https://doi.org/10.2312/vcbm.20181228}{doi:\nolinkurl{10.2312/vcbm.20181228}}


\bibitem[Prince et~al\mbox{.}(2025)]%
        {Prince2025}
\bibfield{author}{\bibinfo{person}{Eric~W. Prince}, \bibinfo{person}{Todd~C. Hankinson}, {and} \bibinfo{person}{Carsten G{\"{o}}rg}.} \bibinfo{year}{2025}\natexlab{}.
\newblock \showarticletitle{{A Visual Analytics Framework for Assessing Interactive AI for Clinical Decision Support}}.
\newblock \bibinfo{journal}{\emph{Pacific Symposium on Biocomputing. Pacific Symposium on Biocomputing}}  \bibinfo{volume}{30} (\bibinfo{year}{2025}), \bibinfo{pages}{40--53}.
\newblock
\showISSN{23356936}
\href{https://doi.org/10.1142/9789819807024{\_}0004}{doi:\nolinkurl{10.1142/9789819807024{\_}0004}}


\bibitem[Pushkarna et~al\mbox{.}(2022)]%
        {Pushkarna2022}
\bibfield{author}{\bibinfo{person}{Mahima Pushkarna}, \bibinfo{person}{Andrew Zaldivar}, {and} \bibinfo{person}{Oddur Kjartansson}.} \bibinfo{year}{2022}\natexlab{}.
\newblock \showarticletitle{{Data Cards: Purposeful and Transparent Dataset Documentation for Responsible AI}}. In \bibinfo{booktitle}{\emph{ACM International Conference Proceeding Series}}. \bibinfo{publisher}{Association for Computing Machinery}, \bibinfo{pages}{1776--1826}.
\newblock
\showISBNx{9781450393522}
\href{https://doi.org/10.1145/3531146.3533231}{doi:\nolinkurl{10.1145/3531146.3533231}}


\bibitem[Raji et~al\mbox{.}(2020)]%
        {Raji2020}
\bibfield{author}{\bibinfo{person}{Inioluwa~Deborah Raji}, \bibinfo{person}{Andrew Smart}, \bibinfo{person}{Rebecca~N. White}, \bibinfo{person}{Margaret Mitchell}, \bibinfo{person}{Timnit Gebru}, \bibinfo{person}{Ben Hutchinson}, \bibinfo{person}{Jamila Smith-Loud}, \bibinfo{person}{Daniel Theron}, {and} \bibinfo{person}{Parker Barnes}.} \bibinfo{year}{2020}\natexlab{}.
\newblock \showarticletitle{{Closing the AI accountability gap: Defining an end-to-end framework for internal algorithmic auditing}}. In \bibinfo{booktitle}{\emph{FAT* 2020 - Proceedings of the 2020 Conference on Fairness, Accountability, and Transparency}}. \bibinfo{publisher}{Association for Computing Machinery, Inc}, \bibinfo{pages}{33--44}.
\newblock
\showISBNx{9781450369367}
\href{https://doi.org/10.1145/3351095.3372873}{doi:\nolinkurl{10.1145/3351095.3372873}}


\bibitem[Rigby(2019)]%
        {Rigby2019}
\bibfield{author}{\bibinfo{person}{Michael~J Rigby}.} \bibinfo{year}{2019}\natexlab{}.
\newblock \bibinfo{booktitle}{\emph{{FROM THE EDITOR Ethical Dimensions of Using Artificial Intelligence in Health Care}}}.
\newblock \bibinfo{type}{{T}echnical {R}eport}~2. \bibinfo{pages}{121--124} pages.
\newblock
\urldef\tempurl%
\url{www.amajournalofethics.org}
\showURL{%
\tempurl}


\bibitem[Roy et~al\mbox{.}(2025)]%
        {Roy2025}
\bibfield{author}{\bibinfo{person}{Arjun Roy}, \bibinfo{person}{Stavroula Rizou}, \bibinfo{person}{Symeon Papadopoulos}, {and} \bibinfo{person}{Eirini Ntoutsi}.} \bibinfo{year}{2025}\natexlab{}.
\newblock \showarticletitle{{Achieving Socio-Economic Parity through the Lens of EU AI Act}}. In \bibinfo{booktitle}{\emph{ACMF AccT 2025 - Proceedings of the 2025 ACM Conference on Fairness, Accountability,and Transparency}}. \bibinfo{publisher}{Association for Computing Machinery, Inc}, \bibinfo{pages}{1890--1901}.
\newblock
\showISBNx{9798400714825}
\href{https://doi.org/10.1145/3715275.3732125}{doi:\nolinkurl{10.1145/3715275.3732125}}


\bibitem[Ruster and Davis(2025)]%
        {Ruster2025}
\bibfield{author}{\bibinfo{person}{Lorenn~P. Ruster} {and} \bibinfo{person}{Jenny~L. Davis}.} \bibinfo{year}{2025}\natexlab{}.
\newblock \showarticletitle{{The Gaps that Never Were: Reconsidering Responsible AI's Principle-Practice Problem}}. In \bibinfo{booktitle}{\emph{ACMF AccT 2025 - Proceedings of the 2025 ACM Conference on Fairness, Accountability,and Transparency}}. \bibinfo{publisher}{Association for Computing Machinery, Inc}, \bibinfo{pages}{350--360}.
\newblock
\showISBNx{9798400714825}
\href{https://doi.org/10.1145/3715275.3732024}{doi:\nolinkurl{10.1145/3715275.3732024}}


\bibitem[Sabuncuoglu et~al\mbox{.}(2025)]%
        {Sabuncuoglu2025}
\bibfield{author}{\bibinfo{person}{Alpay Sabuncuoglu}, \bibinfo{person}{Christopher Burr}, {and} \bibinfo{person}{Carsten Maple}.} \bibinfo{year}{2025}\natexlab{}.
\newblock \showarticletitle{{Justified Evidence Collection for Argument-based AI Fairness Assurance}}. In \bibinfo{booktitle}{\emph{ACMF AccT 2025 - Proceedings of the 2025 ACM Conference on Fairness, Accountability,and Transparency}}. \bibinfo{publisher}{Association for Computing Machinery, Inc}, \bibinfo{pages}{18--28}.
\newblock
\showISBNx{9798400714825}
\href{https://doi.org/10.1145/3715275.3732003}{doi:\nolinkurl{10.1145/3715275.3732003}}


\bibitem[Sadek et~al\mbox{.}(2024)]%
        {Sadek2024}
\bibfield{author}{\bibinfo{person}{Malak Sadek}, \bibinfo{person}{Emma Kallina}, \bibinfo{person}{Thomas Bohn{\'{e}}}, \bibinfo{person}{Céline Mougenot}, \bibinfo{person}{Rafael~A. Calvo}, {and} \bibinfo{person}{Stephen Cave}.} \bibinfo{year}{2024}\natexlab{}.
\newblock \showarticletitle{{Challenges of responsible AI in practice: scoping review and recommended actions}}.
\newblock \bibinfo{journal}{\emph{AI and Society}} (\bibinfo{date}{1} \bibinfo{year}{2024}).
\newblock
\showISSN{14355655}
\href{https://doi.org/10.1007/s00146-024-01880-9}{doi:\nolinkurl{10.1007/s00146-024-01880-9}}


\bibitem[Scheuerman(2024)]%
        {Scheuerman2024}
\bibfield{author}{\bibinfo{person}{Morgan~Klaus Scheuerman}.} \bibinfo{year}{2024}\natexlab{}.
\newblock \showarticletitle{{In the Walled Garden: Challenges and Opportunities for Research on the Practices of the AI Tech Industry}}. In \bibinfo{booktitle}{\emph{2024 ACM Conference on Fairness, Accountability, and Transparency, FAccT 2024}}. \bibinfo{publisher}{Association for Computing Machinery, Inc}, \bibinfo{pages}{456--466}.
\newblock
\showISBNx{9798400704505}
\href{https://doi.org/10.1145/3630106.3658918}{doi:\nolinkurl{10.1145/3630106.3658918}}


\bibitem[Schulze et~al\mbox{.}(2023)]%
        {Schulze2023}
\bibfield{author}{\bibinfo{person}{Annett Schulze}, \bibinfo{person}{Fabian Brand}, \bibinfo{person}{Johanna Geppert}, {and} \bibinfo{person}{Gaby~Fleur B{\"{o}}l}.} \bibinfo{year}{2023}\natexlab{}.
\newblock \bibinfo{title}{{Digital dashboards visualizing public health data: a systematic review}}.
\newblock
\showISSN{22962565}
\href{https://doi.org/10.3389/fpubh.2023.999958}{doi:\nolinkurl{10.3389/fpubh.2023.999958}}


\bibitem[Sedlmair et~al\mbox{.}(2012)]%
        {SMM2012}
\bibfield{author}{\bibinfo{person}{Michael Sedlmair}, \bibinfo{person}{Miriah Meyer}, {and} \bibinfo{person}{Tamara Munzner}.} \bibinfo{year}{2012}\natexlab{}.
\newblock \showarticletitle{{Design Study Methodology: Reflections from the Trenches and the Stacks}}.
\newblock \bibinfo{journal}{\emph{EEE Transactions on Visualization and Computer Graphics}} (\bibinfo{year}{2012}).
\newblock
\showISBNx{10772626/12}


\bibitem[Siala and Wang(2022)]%
        {Siala2022}
\bibfield{author}{\bibinfo{person}{Haytham Siala} {and} \bibinfo{person}{Yichuan Wang}.} \bibinfo{year}{2022}\natexlab{}.
\newblock \showarticletitle{{SHIFTing artificial intelligence to be responsible in healthcare: A systematic review}}.
\newblock \bibinfo{journal}{\emph{Social Science and Medicine}}  \bibinfo{volume}{296} (\bibinfo{date}{3} \bibinfo{year}{2022}).
\newblock
\showISSN{18735347}
\href{https://doi.org/10.1016/j.socscimed.2022.114782}{doi:\nolinkurl{10.1016/j.socscimed.2022.114782}}


\bibitem[Steimers and Schneider(2022)]%
        {Steimers2022}
\bibfield{author}{\bibinfo{person}{André Steimers} {and} \bibinfo{person}{Moritz Schneider}.} \bibinfo{year}{2022}\natexlab{}.
\newblock \showarticletitle{{Sources of Risk of AI Systems}}.
\newblock \bibinfo{journal}{\emph{International Journal of Environmental Research and Public Health}} \bibinfo{volume}{19}, \bibinfo{number}{6} (\bibinfo{date}{3} \bibinfo{year}{2022}).
\newblock
\showISSN{16604601}
\href{https://doi.org/10.3390/ijerph19063641}{doi:\nolinkurl{10.3390/ijerph19063641}}


\bibitem[Surodina and Borgo(2025)]%
        {Surodina2025}
\bibfield{author}{\bibinfo{person}{S Surodina} {and} \bibinfo{person}{R Borgo}.} \bibinfo{year}{2025}\natexlab{}.
\newblock \showarticletitle{{What Makes a Design Study Sustainable in Complex Domains? A Characterisation Framework for Regulated, Stakeholder-Rich Contexts}}.
\newblock  (\bibinfo{year}{2025}).
\newblock
\href{https://doi.org/10.2312/cgvc.20251216}{doi:\nolinkurl{10.2312/cgvc.20251216}}


\bibitem[Surodina et~al\mbox{.}(2024)]%
        {Surodina2024}
\bibfield{author}{\bibinfo{person}{S Surodina}, \bibinfo{person}{D Volkova}, \bibinfo{person}{A Abdul-Rahman}, {and} \bibinfo{person}{R Borgo}.} \bibinfo{year}{2024}\natexlab{}.
\newblock \showarticletitle{{Visualizing Complex Data Decisions: Design Study for Ethical Factors in AI Clinical Decision Support Systems}}.
\newblock \bibinfo{journal}{\emph{Computer Graphics {\&} Visual Computing}} (\bibinfo{year}{2024}).
\newblock
\href{https://doi.org/10.2312/cgvc.20241227}{doi:\nolinkurl{10.2312/cgvc.20241227}}


\bibitem[Turri et~al\mbox{.}(2024)]%
        {Turri2024}
\bibfield{author}{\bibinfo{person}{Violet Turri}, \bibinfo{person}{Katelyn Morrison}, \bibinfo{person}{Katherine~Marie Robinson}, \bibinfo{person}{Collin Abidi}, \bibinfo{person}{Adam Perer}, \bibinfo{person}{Jodi Forlizzi}, {and} \bibinfo{person}{Rachel Dzombak}.} \bibinfo{year}{2024}\natexlab{}.
\newblock \showarticletitle{{Transparency in the Wild: Navigating Transparency in a Deployed AI System to Broaden Need-Finding Approaches}}. In \bibinfo{booktitle}{\emph{2024 ACM Conference on Fairness, Accountability, and Transparency, FAccT 2024}}. \bibinfo{publisher}{Association for Computing Machinery, Inc}, \bibinfo{pages}{1494--1514}.
\newblock
\showISBNx{9798400704505}
\href{https://doi.org/10.1145/3630106.3658985}{doi:\nolinkurl{10.1145/3630106.3658985}}


\bibitem[{United Kingdom Department of Health and Social Care}(2025)]%
        {DoH2025}
\bibfield{author}{\bibinfo{person}{{United Kingdom Department of Health and Social Care}}.} \bibinfo{year}{2025}\natexlab{}.
\newblock \bibinfo{title}{{Fit for the Future: 10 Year Health Plan for England}}.
\newblock
\urldef\tempurl%
\url{https://www.gov.uk/government/publications/10-year-health-plan-for-england-fit-for-the-future}
\showURL{%
\tempurl}


\bibitem[Urman et~al\mbox{.}(2025)]%
        {Urman2025}
\bibfield{author}{\bibinfo{person}{Aleksandra Urman}, \bibinfo{person}{Mykola Makhortykh}, {and} \bibinfo{person}{Aniko Hannak}.} \bibinfo{year}{2025}\natexlab{}.
\newblock \showarticletitle{{WEIRD Audits? Research Trends, Linguistic and Geographical Disparities in the Algorithm Audits of Online Platforms-A Systematic Literature Review}}. In \bibinfo{booktitle}{\emph{ACMF AccT 2025 - Proceedings of the 2025 ACM Conference on Fairness, Accountability,and Transparency}}. \bibinfo{publisher}{Association for Computing Machinery, Inc}, \bibinfo{pages}{375--390}.
\newblock
\showISBNx{9798400714825}
\href{https://doi.org/10.1145/3715275.3732026}{doi:\nolinkurl{10.1145/3715275.3732026}}


\bibitem[Valdivia(2025)]%
        {Valdivia2025}
\bibfield{author}{\bibinfo{person}{Ana Valdivia}.} \bibinfo{year}{2025}\natexlab{}.
\newblock \showarticletitle{{Data Ecofeminism}}. In \bibinfo{booktitle}{\emph{ACMF AccT 2025 - Proceedings of the 2025 ACM Conference on Fairness, Accountability,and Transparency}}. \bibinfo{publisher}{Association for Computing Machinery, Inc}, \bibinfo{pages}{391--403}.
\newblock
\showISBNx{9798400714825}
\href{https://doi.org/10.1145/3715275.3732027}{doi:\nolinkurl{10.1145/3715275.3732027}}


\bibitem[Vethman et~al\mbox{.}(2025)]%
        {Vethman2025}
\bibfield{author}{\bibinfo{person}{Steven Vethman}, \bibinfo{person}{Quirine~T.S. Smit}, \bibinfo{person}{Nina~M. Van~Liebergen}, {and} \bibinfo{person}{Cor~J. Veenman}.} \bibinfo{year}{2025}\natexlab{}.
\newblock \showarticletitle{{Fairness beyond the Algorithmic Frame: Actionable Recommendations for an Intersectional Approach}}. In \bibinfo{booktitle}{\emph{ACMF AccT 2025 - Proceedings of the 2025 ACM Conference on Fairness, Accountability,and Transparency}}. \bibinfo{publisher}{Association for Computing Machinery, Inc}, \bibinfo{pages}{3276--3290}.
\newblock
\showISBNx{9798400714825}
\href{https://doi.org/10.1145/3715275.3732210}{doi:\nolinkurl{10.1145/3715275.3732210}}


\bibitem[Wang et~al\mbox{.}(2023)]%
        {Wang2023}
\bibfield{author}{\bibinfo{person}{Liuping Wang}, \bibinfo{person}{Zhan Zhang}, \bibinfo{person}{Dakuo Wang}, \bibinfo{person}{Weidan Cao}, \bibinfo{person}{Xiaomu Zhou}, \bibinfo{person}{Ping Zhang}, \bibinfo{person}{Jianxing Liu}, \bibinfo{person}{Xiangmin Fan}, {and} \bibinfo{person}{Feng Tian}.} \bibinfo{year}{2023}\natexlab{}.
\newblock \showarticletitle{{Human-centered design and evaluation of AI-empowered clinical decision support systems: a systematic review}}.
\newblock \bibinfo{journal}{\emph{Frontiers in Computer Science}}  \bibinfo{volume}{5} (\bibinfo{year}{2023}).
\newblock
\showISSN{26249898}
\href{https://doi.org/10.3389/fcomp.2023.1187299}{doi:\nolinkurl{10.3389/fcomp.2023.1187299}}


\bibitem[Wu et~al\mbox{.}(2023)]%
        {Wu2023}
\bibfield{author}{\bibinfo{person}{Aoyu Wu}, \bibinfo{person}{Dazhen Deng}, \bibinfo{person}{Min Chen}, \bibinfo{person}{Shixia Liu}, \bibinfo{person}{Daniel Keim}, \bibinfo{person}{Ross MacIejewski}, \bibinfo{person}{Silvia Miksch}, \bibinfo{person}{Hendrik Strobelt}, \bibinfo{person}{Fernanda Viegas}, {and} \bibinfo{person}{Martin Wattenberg}.} \bibinfo{year}{2023}\natexlab{}.
\newblock \showarticletitle{{Grand Challenges in Visual Analytics Applications}}.
\newblock \bibinfo{journal}{\emph{IEEE Computer Graphics and Applications}} \bibinfo{volume}{43}, \bibinfo{number}{5} (\bibinfo{date}{9} \bibinfo{year}{2023}), \bibinfo{pages}{83--90}.
\newblock
\showISSN{15581756}
\href{https://doi.org/10.1109/MCG.2023.3284620}{doi:\nolinkurl{10.1109/MCG.2023.3284620}}


\bibitem[Xing et~al\mbox{.}(2024)]%
        {Xing2024}
\bibfield{author}{\bibinfo{person}{Yiwen Xing}, \bibinfo{person}{Gabriel~D. Cantareira}, \bibinfo{person}{Rita Borgo}, {and} \bibinfo{person}{Alfie Abdul-Rahman}.} \bibinfo{year}{2024}\natexlab{}.
\newblock \showarticletitle{{A Review and Analysis of Evaluation Practices in VIS Domain Applications}}.
\newblock \bibinfo{journal}{\emph{IEEE Transactions on Visualization and Computer Graphics}} (\bibinfo{year}{2024}).
\newblock
\showISSN{19410506}
\href{https://doi.org/10.1109/TVCG.2024.3460181}{doi:\nolinkurl{10.1109/TVCG.2024.3460181}}


\bibitem[Ye et~al\mbox{.}(2020)]%
        {Ye2020}
\bibfield{author}{\bibinfo{person}{Yucong~Chris Ye}, \bibinfo{person}{Franz Sauer}, \bibinfo{person}{Kwan~Liu Ma}, \bibinfo{person}{Konduri Aditya}, {and} \bibinfo{person}{Jacqueline Chen}.} \bibinfo{year}{2020}\natexlab{}.
\newblock \showarticletitle{{A User-Centered Design Study in Scientific Visualization Targeting Domain Experts}}.
\newblock \bibinfo{journal}{\emph{IEEE Transactions on Visualization and Computer Graphics}} \bibinfo{volume}{26}, \bibinfo{number}{6} (\bibinfo{date}{6} \bibinfo{year}{2020}), \bibinfo{pages}{2192--2203}.
\newblock
\showISSN{19410506}
\href{https://doi.org/10.1109/TVCG.2020.2970525}{doi:\nolinkurl{10.1109/TVCG.2020.2970525}}


\bibitem[Yew et~al\mbox{.}(2025)]%
        {Yew2025}
\bibfield{author}{\bibinfo{person}{Rui~Jie Yew}, \bibinfo{person}{Bill Marino}, {and} \bibinfo{person}{Suresh Venkatasubramanian}.} \bibinfo{year}{2025}\natexlab{}.
\newblock \showarticletitle{{Red Teaming AI Policy: A Taxonomy of Avoision and the EU AI Act}}. In \bibinfo{booktitle}{\emph{ACMF AccT 2025 - Proceedings of the 2025 ACM Conference on Fairness, Accountability,and Transparency}}. \bibinfo{publisher}{Association for Computing Machinery, Inc}, \bibinfo{pages}{404--415}.
\newblock
\showISBNx{9798400714825}
\href{https://doi.org/10.1145/3715275.3732028}{doi:\nolinkurl{10.1145/3715275.3732028}}


\bibitem[Zhuang et~al\mbox{.}(2022)]%
        {Zhuang2022}
\bibfield{author}{\bibinfo{person}{Mengdie Zhuang}, \bibinfo{person}{David Concannon}, {and} \bibinfo{person}{Ed Manley}.} \bibinfo{year}{2022}\natexlab{}.
\newblock \showarticletitle{{A Framework for Evaluating Dashboards in Healthcare}}.
\newblock \bibinfo{journal}{\emph{IEEE Transactions on Visualization and Computer Graphics}} \bibinfo{volume}{28}, \bibinfo{number}{4} (\bibinfo{date}{4} \bibinfo{year}{2022}), \bibinfo{pages}{1715--1731}.
\newblock
\showISSN{1077-2626}
\href{https://doi.org/10.1109/TVCG.2022.3147154}{doi:\nolinkurl{10.1109/TVCG.2022.3147154}}


\end{thebibliography}

\appendix

\end{document}